\documentclass[12pt]{article} 

\RequirePackage[T1]{fontenc}

\usepackage[height=8.85in,width=6.45in]{geometry}
\usepackage{amsmath}
\usepackage{amssymb}
\usepackage{slashed}
\usepackage{braket}
\usepackage[svgnames]{xcolor}
\usepackage[colorlinks,citecolor=DarkGreen,linkcolor=FireBrick]{hyperref}
\usepackage{cite}
\usepackage{tikz}
\usepackage{tikz-cd}
\usepackage{times}
\usepackage{courier}
\usepackage{bm}
\usepackage{mdframed}
\usepackage{datetime}
\usepackage{dsfont}
\usepackage{multirow}
\usepackage{cancel}
\usepackage{caption}
\usepackage{subcaption}
\usepackage{graphicx}
\usepackage{pifont}
\usepackage{upgreek}
\usepackage{enumitem}
\usepackage{cleveref}

\allowdisplaybreaks
\numberwithin{equation}{section}
\graphicspath{{./figures/}}

\newcommand{\Res}{\textrm{Res}}

\renewcommand{\arraystretch}{1.5}

\begin{document}
\begin{titlepage}

\begin{flushright}
UNIST-MTH-24-RS-01
\end{flushright}

\vskip 3cm

\begin{center}	
	{\Large \bfseries 
	    The Origin of Calabi-Yau Crystals in BPS States Counting
	}	
	\vskip 1cm
           Jiakang Bao$^{1}$,
           Rak-Kyeong Seong$^{2}$,
           Masahito Yamazaki$^{1,3}$	
	\vskip 1cm

 \def\arraystretch{1}
	\begin{tabular}{c}
		$^1$ Kavli Institute for the Physics and Mathematics of the Universe,\\
		    University of Tokyo, Kashiwa, Chiba 277-8583, Japan\\
		$^2$ Department of Mathematical Sciences and Department of Physics,\\
            Ulsan National Institute of Science and Technology, 50 UNIST-gil, Ulsan 44919, South Korea\\
	    $^3$ Trans-Scale Quantum Science Institute, 
		        University of Tokyo, Tokyo 113-0033, Japan\\
	\end{tabular}

 \vskip 0.5cm
 \href{malto:jiakang.bao@ipmu.jp}{jiakang.bao@ipmu.jp},
 \href{malto:seong@unist.ac.kr}{seong@unist.ac.kr},
 \href{malto:masahito.yamazaki@ipmu.jp}{masahito.yamazaki@ipmu.jp}
 \vskip 1cm
	
\end{center}

\noindent
We study the counting problem of BPS D-branes wrapping holomorphic cycles of a general toric Calabi-Yau manifold. 
We evaluate the Jeffrey-Kirwan residues for the flavoured Witten index for the supersymmetric quiver quantum mechanics on the worldvolume of the D-branes, and find that 
BPS degeneracies are described by a statistical mechanical model of crystal melting. For Calabi-Yau threefolds, we reproduce the crystal melting models long known in the literature. For Calabi-Yau fourfolds, however, 
we find that the crystal does not contain the full information for the BPS degeneracy and we need to explicitly evaluate non-trivial weights assigned to the crystal configurations. Our discussions
treat Calabi-Yau threefolds and fourfolds on equal footing, 
and include discussions on elliptic and rational generalizations of the BPS states counting, connections to the mathematical definition of generalized Donaldson-Thomas invariants, examples of wall crossings, and of trialities in quiver gauge theories.
\end{titlepage}

\setcounter{tocdepth}{3}
\tableofcontents

\section{Introduction and Summary}\label{intro}
One of the fascinating aspects of supersymmetry is that it allows one to obtain non-perturbative results which are hard to be obtained otherwise. With the help of the supersymmetric localization (see e.g.\ \cite{Pestun:2016zxk} for a review), the infinite-dimensional path integral can be reduced to a finite-dimensional integral of over the moduli space of Bogomolnyi-Prasad-Sommerfield (BPS) configurations, so that 
we can exactly compute partition functions and physical observables of the theory. This has led to precision studies of supersymmetric gauge theories, black holes and string theory.

In this paper, we are interested in the counting problem of the supersymmetric ground states of Type IIA string theory on a Calabi-Yau (CY) manifold, whose holomorphic cycles are wrapped by D$p$-brane with even $p$'s. In particular we consider non-compact toric Calabi-Yau threefolds and fourfolds.

On the one hand, 
on the worldvolume of the D-branes we find supersymmetric quiver quantum mechanics. We expect that the BPS degeneracy can be computed by 
evaluating the BPS index (Witten index \cite{Witten:1982df}) of the quantum mechanics, with  fugacities turned on for  global symmetries.

On the other hand, when putting Type IIA string theory on a toric Calabi-Yau threefold, the counting problem of BPS states can be nicely translated into the combinatorial problem of crystal melting. The statistical mechanical model of crystal melting for a general toric CY threefold was formulated in 
\cite{Ooguri:2008yb} (see also \cite{Szendroi:2007nu,mozgovoy2010noncommutative,Jafferis:2008uf,Chuang:2008aw,Ooguri:2009ri,Dimofte:2009bv,Nagao:2009rq,Sulkowski:2009rw,Aganagic:2009cg,Ooguri:2010yk,Nishinaka:2010fh,Nishinaka:2010qk,Cirafici:2010bd,Yamazaki:2011wy}, and \cite{Yamazaki:2010fz} for a review), generalizing the crystal melting model for $\mathbb{C}^3$ \cite{Okounkov:2003sp}; the BPS degeneracies can be obtained by enumerating all the possible configurations of the molten crystals satisfying the melting rule. Mathematically, these BPS invariants are in fact the generalized Donaldson-Thomas (DT) invariants \cite{Thomas:1998uj}, 
and the crystal configurations are nothing but the fixed-point sets of the moduli space, in the equivariant localization with respect to the torus action originating from the toric geometry.

One of the main goals of this paper is to connect these two descriptions, by directly deriving the crystal melting model from the 
evaluation of supersymmetric indices by
applying the aforementioned supersymmetric localization techniques. 
Following \cite{Benini:2013xpa}, the integral of the 1-loop determinant can be computed using the Jeffrey-Kirwan (JK) residue \cite{MR1318878}. For the cases of the toric CY threefolds, we discuss them in Appendix \ref{3dcrystals}. It is worth noting that recently the relations between JK residues and DT$_3$ invariants were proven mathematically in \cite{Ontani:2021nia,ontani2023virtual} (see also \cite{Beaujard:2019pkn,Mozgovoy:2020has,Descombes:2021snc} for relevant calculations). Moreover, our discussion generalizes (with interesting new features) to toric CY fourfolds. We shall consider the BPS bound states formed by D6-/D4-/D2-/D0-branes wrapping compact cycles in the CY fourfold, with a pair of non-compact D8- and anti-D8-branes filling the CY.

There are clear similarities between the BPS state countings on threefolds and fourfolds.
For example, the definition of the 3d crystals can be naturally extended to those of the 4d crystals, and it is natural to expect that the 4d crystals would likewise give a combinatorial interpretation of the BPS spectra. Indeed, thanks to the JK residue formula, we can actually show that the torus fixed points are in one-to-one correspondence with the crystal configurations, for both the threefolds and the fourfolds. The melting rule of the crystals is then a natural consequence of the pole structure in the JK residue formula.
As we are considering the BPS indices, there would also be signs given by the fermion number. For CY threefolds, we shall not only obtain the expected crystal structure, but also recover the correct signs from the JK residues.

For toric CY fourfolds, the 4d crystals were recently discussed in \cite{Franco:2023tly}, where the combinatorial structure can be nicely obtained from the periodic quivers and brane brick models. Here, we would like to discuss how the 4d crystals could appear from a different perspective, namely the BPS index computations, where the situations for the toric CY fourfolds could be more complicated compared to the threefolds.
Although we still have the 4d crystals, it is important to emphasize that the crystals themselves are \textit{not} sufficient for BPS counting. Although the 4d crystals still label the isolated fixed points of the moduli space, they cannot encode the full information of the BPS states; the full information 
can be extracted only when supplemented by the weights obtained from the JK residue formula, and the weights are rational functions of the fugacities associated to the global symmetries. This is a significant difference from the threefold cases.

Let us discuss the fourfold cases in more detail. In the 1-loop determinant for the index, there are contributions from chiral and Fermi multiplets, as well as from vector multiplets. The information of the supermultiplets is nicely encoded by a 2d $\mathcal{N}=(0,2)$ quiver.
The corresponding 2d $\mathcal{N}=(0,2)$ quiver gauge theory can be considered as the worldvolume theory of probe D1-branes on the Calabi-Yau fourfold. This class of theories is represented by a Type IIA brane configuration known as a brane brick model \cite{Franco:2015tna,Franco:2015tya,Franco:2016nwv,Franco:2016qxh,Franco:2016fxm,Franco:2017cjj,Franco:2018qsc,Franco:2022gvl,Franco:2022isw,Kho:2023dcm,Franco:2023tyf}.
The dimensional reduction of this class of theories gives rise to an $\mathcal{N}=2$ supersymmetric quiver quantum mechanics. The non-compact D8-/anti-D8-brane pair corresponds to the ``framing'' of the quiver, which represents the non-dynamical flavour branes. In particular, the chiral and Fermi multiplets thereof come from the strings connected to the D8 and anti-D8 respectively.

We also turn on a background B-field \cite{Witten:2000mf}, and the chamber structures could depend on its value. In certain limit of the fugacities/chemical potentials, we expect the tachyon condensation to happen, and the D8/anti-D8 pair would annihilate into a single D6-brane. For $X\times\mathbb{C}$ where $X$ is a toric CY threefold, this should then recover the partition function for $X$. From the partition functions we have for $\mathbb{C}^2\times\mathbb{C}^2/\mathbb{Z}_n$ and conifold$\times\mathbb{C}$, we find that the limit should be $\epsilon_k/(v_1-v_2)\rightarrow0$ (for any $k=1,2,3,4$), where $\epsilon_k$ and $v_{1,2}$ are the equivariant parameters associated to the CY isometries and the framing respectively.

Moreover, as we are considering the D8/anti-D8 pair (though considering D$p$-branes bound to a single D8 is still well-posed), the framing node should be U$(1|1)$ (instead of U$(1)$). In terms of the crystal structure that labels the fixed points, it is reflected by the fact that when $v_2$ is tuned to be certain linear combinations of $v_1$ and $\epsilon_k$, the crystal would get truncated. Indeed, in the JK residue formula, this may cause extra cancellations of the factors in the numerator and the denominator in the 1-loop determinant, and thus terminates the growth of the crystal at the corresponding atom(s).

As an illustration, we shall discuss some examples of 2d $\mathcal{N}=(0,2)$ theories given by brane brick models and compute their BPS partition functions. As the stability of the BPS states could vary for different moduli, there is also the wall crossing phenomenon. We shall consider the chambers that can be reached via ``mutations'' of the framed quivers. For 2d $\mathcal{N}=(0,2)$ quivers that can be obtained from dimensional reduction of 4d $\mathcal{N}=1$ quivers, some of the chamber structures can be naturally inherited from the threefold counterparts. On the other hand, for 2d $\mathcal{N}=(0,2)$ theories themselves, they also enjoy certain IR equivalence known as the triality \cite{Gadde:2013lxa} (see also \cite{Closset:2017yte}) that were shown to have a natural interpretation in terms of brane brick models \cite{Franco:2016nwv}. Therefore, it is expected that there is a richer chamber structure under wall crossing for the fourfold cases. It would be an interesting problem to systematically extend this discussion to more general examples of toric Calabi-Yau fourfolds.

In mathematics literature, there are also extensive studies on defining the (generalized) DT invariants for the CY fourfolds. For instance, such invariants were introduced and explored in \cite{cao2014donaldson,Cao:2017swr,Cao:2019tnw,Cao:2020vce,Cao:2019fqq,Cao:2019tvv} using the obstruction theory. It turns out that they have some nice physical interpretations, where certain Ext groups correspond to the multiplets in the gauge theories and the insertions are related to the framings. Moreover, the DT invariant in this mathematical definition depends on the choice of the orientation of a certain real line bundle on the Hilbert scheme. We will see that this is a choice of the sign collectively from the $J$- and $E$-terms of the gauge theory in each crystal configuration, and would be canonically determined with the physical input.

Let us also mention that for the cases of Calabi-Yau threefolds, it has recently been found that there exist some infinite-dimensional algebras,
known as the (shifted) quiver Yangians \cite{Li:2020rij,Galakhov:2020vyb,Galakhov:2021xum,Noshita:2021ldl,Galakhov:2021vbo,Yamazaki:2022cdg}, as the BPS state algebras underlying the BPS state counting---the 3d crystals are the weight spaces of the 
representations of the quiver Yangians, and the BPS partition functions are nothing but the characters of the quiver Yangians for the crystal representations. It is natural to imagine that similar algebras exist for the cases of Calabi-Yau fourfolds, which have the 
4d crystals as the representation spaces\footnote{The charge function introduced in \cite{Galakhov:2023vic} could play a role in such representations.}. Let us again emphasize that, however, the situation for Calabi-Yau fourfolds is more complicated since the 4d crystal in itself is not sufficient to fully recover the data of the physical BPS state counting. The BPS algbera should also incorporate the non-trivial weights as mentioned above.

The paper is organized as follows. In \S\ref{review_1}, we will review 2d $\mathcal{N}=(0,2)$ quiver gauge theories associated to toric CY fourfolds and their realization in terms of brane brick models. In \S\ref{BPS}, after recalling the JK residue formula used for localization, we will derive the combinatorial part, namely the 4d crystals, in the BPS counting problem for the fourfolds. We shall also comment on the extra data of the BPS states that are not encoded by the crystals, as well as the wall crossing phenomenon. In \S\ref{ellipticandrational}, we will mention the elliptic and rational counterparts of the partition functions, where the elliptic invariants have further constraints on the parameters due to the anomalies. Some explicit examples will be given in \S\ref{ex}. In \S\ref{DT4}, we will discuss the connections between the BPS counting and the mathematical definition of the DT invariants. In Appendix \ref{ellg}, we list the contributions from the supermultiplets in the integrands for the elliptic genera for both 2d $\mathcal{N}=(2,2)$ and $\mathcal{N}=(0,2)$ theories. We will show in Appendix \ref{3dcrystals} how the 3d crystals, as well as the correct signs in the BPS indices, can be obtained from the JK residue formula for toric CY threefolds.

\section{\texorpdfstring{Toric CY$_4$ and 2d $\mathcal{N}=(0,2)$ Theories}{Toric CY4 and 2d N=(0,2) Theories}}\label{review_1}

In this paper, we study an $\mathcal{N}=2$ supersymmetric quiver quantum 
mechanics associated with a toric Calabi-Yau fourfold compactifications of 
Type IIA string theory. Such a quiver quantum mechanics (with $\mathcal{N}=2$ supercharges) can be 
obtained as the dimensional reduction of a class of 2d $\mathcal{N}=(0,2)$ quiver gauge theories.
These theories are worldvolume theories of D1-branes probing toric Calabi-Yau fourfolds and are realized in terms of a Type IIB brane configuration known as a brane brick model \cite{Franco:2015tna,Franco:2015tya,Franco:2016nwv,Franco:2016qxh,Franco:2016fxm,Franco:2017cjj,Franco:2018qsc,Franco:2022gvl,Franco:2022isw,Kho:2023dcm,Franco:2023tyf}.
The following section gives a brief review of brane brick models. 

\begin{figure}
\centering
    \includegraphics[trim=0 0 0 0,width=3cm]{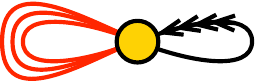}
    \caption{Quiver diagram for the $\mathbb{C}^4$ brane brick model.\label{fquiverex}}
\end{figure}

\paragraph{Quivers.} 
Given a quiver $Q=\{Q_0,Q_1\}$, where the set of vertices is denoted by $Q_0$ and the set of edges by $Q_1$, we can use the following dictionary in order to identify the gauge symmetry and matter content of the corresponding 2d $\mathcal{N}=(0,2)$ gauge theory:
\begin{itemize}
    \item a vertex $a$ represents a U$(N_a)$ gauge group;
    \item an oriented edge from an initial vertex $a$ to a terminal vertex $b$ represents a bifundamental chiral multiplet $X_{ab}$;
    \item an unoriented edge between vertices $a$ and $b$ represents a Fermi multiplet $\Lambda_{ab}$.
\end{itemize}
We note that Fermi fields are not assigned an orientation in the quiver diagram due to the $\Lambda_{ab}\leftrightarrow \overline{\Lambda}_{ab}$ symmetry of 2d $\mathcal{N}=(0,2)$ theories.
Fig.~\ref{fquiverex} shows as an example the quiver diagram for the $\mathbb{C}^4$ brane brick model.

\paragraph{The $J$-and $E$-terms.} 
 Given a Fermi multiplet $\Lambda_{ab}$, there is an associated pair of holomorphic functions of chiral fields called $E_{ab}$ and $J_{ba}$ and are the $J$- and $E$-term relations of the corresponding 2d $\mathcal{N}=(0,2)$ theory. 
 These are restricted to be binomials for toric Calabi-Yau fourfolds, where $E_{ab}$ transforms in the same representation as $\Lambda_{ab}$, while $J_{ba}$ transforms in the conjugate representation of $\Lambda_{ab}$. They satisfy an overall constraint $\sum\text{tr}(E_{ab}J_{ba})=0$. 
The general form of the $J$- and $E$-terms is as follows,
\begin{equation}
    J_{ba}=J_{ba}^+-J_{ba}^-\;,
    \quad 
    E_{ab}=E_{ab}^+-E_{ab}^-\;,
    \label{toricJE}
\end{equation}
where $J_{ba}^{\pm}$ and $E_{ab}^{\pm}$ are monomials in chiral fields. 

As an example, the above $\mathbb{C}^4$ quiver has the following $J$- and $E$-terms:
\begin{equation}
\begin{tabular}{ccc}
 & $J$ & $E$ \\
$\Lambda^{(1)}$: & $YZ-ZY=0$\;, & $DX-XD=0$\;, \\
$\Lambda^{(2)}$: & $ZX-XZ=0$\;, & $DY-YD=0$\;, \\
$\Lambda^{(3)}$: & $XY-YX=0$\;, & $DZ-ZD=0$\;,
\end{tabular}
\end{equation}
where $X,Y,Z,D$ are the chirals and $\Lambda^{(i)}$ correspond to the Fermis.

\begin{figure}
        \centering
        \includegraphics[width=4cm]{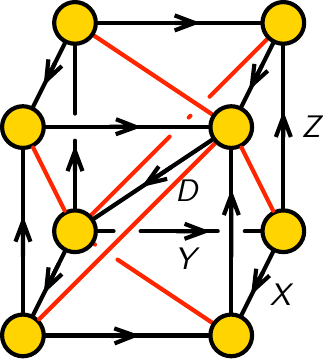}
        \caption{The periodic quiver for the  $\mathbb{C}^4$ brane brick model.\label{periodicquiverex}}
\end{figure}

\paragraph{Periodic Quivers.} The quiver and $J$- and $E$-terms of a 2d $\mathcal{N}=(0,2)$ theory corresponding to a toric Calabi-Yau fourfold can be turned into a periodic quiver on a 3-torus. 
Such a periodic quiver is dual to the underlying Type IIA brane configuration known as a brane brick model that realizes this class of 2d $\mathcal{N}=(0,2)$ theories.
As we will see later, when computing the BPS index, the fixed points of the torus action on the moduli space are actually labeled by a 4-dimensional uplift of the periodic quiver. 
The periodic quiver for the $\mathbb{C}^4$ theory is illustrated in Fig.~\ref{periodicquiverex}.

\begin{figure}
\centering
    \includegraphics[width=13cm]{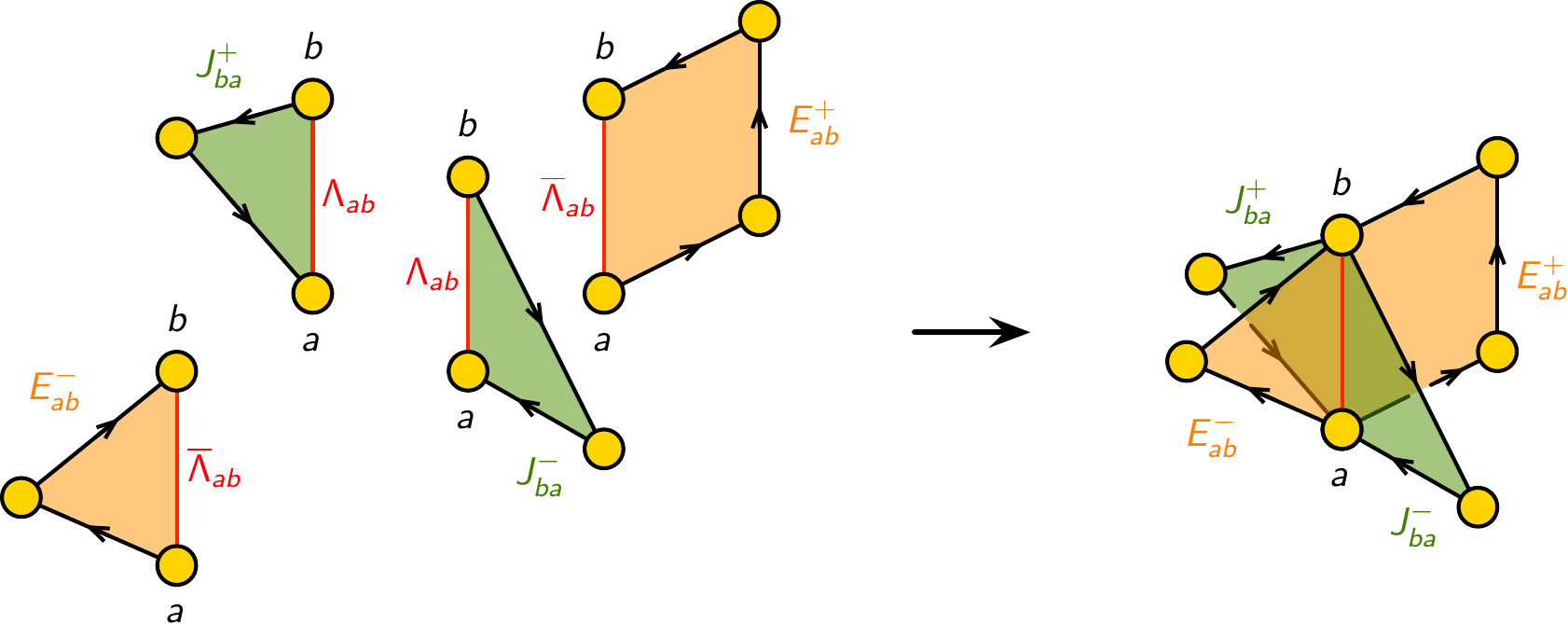}
    \caption{Illustration of $J$- and $E$-term plaquettes that correspond to a single Fermi field in the brane brick model.\label{fplaquettes}}
\end{figure}

In the periodic quiver, each monomial in the $J$- and $E$-terms is represented by a (minimal) plaquette. By plaquette, we mean a closed loop in the periodic quiver composed of multiple chirals and one single Fermi. In particular, the chirals form an oriented path with the two endpoints connected by the Fermi. The toric condition \eqref{toricJE} indicates that there are four plaquettes for each unoriented edge, corresponding to $\left(\Lambda_{ab},J_{ba}^{\pm}\right)$ and $\left(\overline{\Lambda}_{ab},E_{ab}^{\pm}\right)$. Pictorially, this is illustrated in Fig.~\ref{fplaquettes}.

\paragraph{Brane Brick Models.}
Brane brick models \cite{Franco:2015tna,Franco:2015tya,Franco:2016nwv,Franco:2016qxh,Franco:2016fxm,Franco:2017cjj,Franco:2018qsc,Franco:2022gvl,Franco:2022isw,Kho:2023dcm,Franco:2023tyf} represent a large class of 2d $\mathcal{N}=(0,2)$ theories that are realized as worldvolume theories of D1-branes probing toric Calabi-Yau fourfolds.
They represent Type IIA brane configurations of D4-branes suspended from an NS5-brane, forming a tessellation of a 3-torus. 
The periodic quiver of the corresponding 2d $\mathcal{N}=(0,2)$ theory forms the dual graph of this tessellation. 
The brane bricks in the brane brick model correspond to the U$(N)$ gauge groups and brick faces correspond to either chiral or Fermi fields of the corresponding 2d $\mathcal{N}=(0,2)$ theory.

Brane brick models exhibit brick matchings, which are collections of chiral and Fermi fields that cover every plaquette of the brane brick model exactly once. Such brick matchings correspond to GLSM fields in the GLSM description of the moduli spaces associated to abelian brane brick models. 
They are directly related to vertices in the toric diagram of the corresponding toric Calabi-Yau fourfold and 
allow us to better the moduli spaces associated to brane brick models \cite{Franco:2015tna,Franco:2015tya,Kho:2023dcm}.

\paragraph{Dimensional Reduction.} 
The class of 4d $\mathcal{N}=1$ theories that are worldvolume theories of D3-branes probing toric Calabi-Yau threefolds $X$
can be represented by a bipartite periodic graph on a 2-torus known as a brane tiling \cite{Franco:2005rj,Franco:2005sm,Kennaway:2007tq,Yamazaki:2008bt}.
Similar to brick matchings in brane brick models, brane tilings exhibit special collections of chiral fields in the 4d theory that are given by perfect matchings in the associated bipartite graph. 
These perfect matchings correspond to GLSM fields in the toric description of the corresponding Calabi-Yau threefold $X$.

When one dimensionally reduces a 4d $\mathcal{N}=1$ theory given by a brane tiling, the resulting 2d $\mathcal{N}=(2,2)$ theory is known to correspond to a brane brick model associated to a toric Calabi-Yau fourfold of the form $X\times \mathbb{C}$ \cite{Franco:2015tna,Franco:2015tya,Franco:2016fxm}.
The 4d vector $\mathcal{V}_a$ and 4d chiral multiplets $\mathcal{X}_{ab}$ become under dimensional reduction 2d $\mathcal{N}=(2,2)$ vector and chiral multiplets, which in turn 
can be represented in terms of 2d $\mathcal{N}=(0,2)$ multiplets as follows:
\begin{itemize}
    \item \underline{4d $\mathcal{N}=1$ vector $\mathcal{V}_a$:} 2d $\mathcal{N}=(0,2)$ vector $V_a$ $+$ 2d $\mathcal{N}=(0,2)$ adjoint chiral $\Phi_{aa}$;
    \item \underline{4d $\mathcal{N}=1$ chiral $\mathcal{X}_{ab}$:} 2d $\mathcal{N}=(0,2)$ chiral $X_{ab}$ $+$ 2d $\mathcal{N}=(0,2)$ Fermi $\Lambda_{ab}$.
\end{itemize}
The superpotential $W$ of the 4d $\mathcal{N}=1$ gives rise to the $J$- and $E$-terms of the 2d theory as follows,
\begin{equation}
J_{ba} = \frac{\partial W}{\partial X_{ab}} ~,~
E_{ab} = \Phi_{aa} X_{ab} + X_{ab} \Phi_{bb} ~,~
\end{equation}
for every chiral $X_{ab}$ and Fermi $\Lambda_{ab}$ coming from a 4d $\mathcal{N}=1$ chiral $\mathcal{X}_{ab}$.

We note that dimensional reduction of brane tilings into brane brick models can be generalized with an additional orbifolding that breaks the factorization of the toric Calabi-Yau fourfold $X\times \mathbb{C}$.
This process is known as orbifold reduction in the literature \cite{Franco:2016fxm}.
A further generalization of this process is known as 3d printing of brane brick models \cite{Franco:2018qsc}.

\paragraph{Triality.} 
It is known that 2d $\mathcal{N}=(0,2)$ theories exhibit IR dualities similar to Seiberg duality \cite{Seiberg:1994pq} for 4d $\mathcal{N}=1$ theories.
This IR phenomenon for 2d $\mathcal{N}=(0,2)$ theories is known as triality \cite{Gadde:2013lxa,Closset:2017yte}. As the name suggests, 
this low energy equivalence between 2d $\mathcal{N}=(0,2)$ theories
leads to the original theory after performing the triality transformation three times on the same gauge group of the 2d theory.
Triality has a natural interpretation in terms of a local mutation of the corresponding brane brick model \cite{Franco:2016nwv}.
When one performs the local mutation three times, triality returns the original brane brick model.
As will be discussed in \S\ref{wallcrossing}, triality for brane brick models is closely related to the wall crossing phenomenon in the BPS state counting problem.

\section{The 4d Crystals from BPS States Countings}\label{BPS}
In this section, we shall use the JK residue formula to write the index. We will use this to obtain the crystals for BPS counting, together with the non-trivial weights that are not incorporated in the crystals.

\subsection{JK Residues for Supersymmetric Indices}\label{index}

Let us quickly summarize the JK residue formula for the supersymmetric index \cite{Hori:2014tda,Cordova:2014oxa,Hwang:2014uwa}. The flavoured supersymmetric index is defined as
\begin{align}
\mathcal{I} (y, \{ w_i \}, \{q_i \}) = \textrm{Tr} \left[ (-1)^F \text{e}^{-\beta \{Q, Q^{\dagger}\}} y^R \prod_i w_i^{f_i} \prod_k q_k^{f_k} \right]\;,
\end{align}
where $F$ is the fermion number, and $y, w_i, q_k$ are the fugacities for the R-symmetry and flavour symmetries. In particular, $w_i$ (resp.~$q_k$) will be used to denote the fugacities associated to the framing (resp.~U$(1)$ isometries) for the quiver gauge theories. It would be convenient to write
\begin{equation}
    x_i=\text{e}^{2\pi\text{i}u_i}\;,\quad y=\text{e}^{2\pi\text{i}z}\;,\quad w_i=\text{e}^{2\pi\text{i}v_i}\;,\quad q_k=\text{e}^{2\pi\text{i}\epsilon_k}\;.
\end{equation}
Henceforth, we shall use the variables for the fugacities and the chemical potentials interchangeably. In the case of toric CY fourfolds, we have $(\epsilon_1,\epsilon_2,\epsilon_3,\epsilon_4)\in\text{U}(1)^3$ for the $\Omega$-background. It would also be convenient to introduce
\begin{equation}
    \epsilon=\epsilon_1+\epsilon_2+\epsilon_3+\epsilon_4\;,
    \quad q=q_1q_2q_3q_4\;.
\end{equation}
The CY condition is then $\epsilon=0$, or equivalently, $q=1$, for the (unrefined) indices.

From \cite{Benini:2013xpa}, we learn that the index can be computed using the JK residues:
\begin{equation}
 \mathcal{I}=\frac{1}{|W|}\sum_{u^*\in\mathfrak{M}^*_\text{sing}}\textrm{JK-Res}_{u=u^*}(\bm{Q}(u^*),\eta)\, Z_{\textrm{1-loop}}(y,u) \;,
\end{equation}
where $W$ is the Weyl group, and we have collectively denote $\{u_i\}$ as $u$. The 1-loop determinant $Z_{\textrm{1-loop}}$ denotes the integrand of the residue, and JK-Res denotes the Jeffrey-Kirwan residue. Let us explain each ingredient in turn.

\paragraph{The integrand: $Z_{\textrm{1-loop}}$} The integrand $Z_\text{1-loop}$ factorizes into the contributions from the
$\mathcal{N}=(0,2)$ vector/chiral/Fermi multiplet contributions:
\begin{align}
Z_{\textrm{1-loop}}(u) = \prod_VZ_V(y,u) \prod_{\chi}Z_{\chi}(y,u) \prod_{\Lambda}Z_{\Lambda}(y,u) \;.
\end{align}
The contributions from the different multiplets are given as follows:
\begin{itemize}[leftmargin=*]
    \item Vector multiplet $V$ with gauge group $G$:
    \begin{equation}
        Z_V=\prod_{\alpha\in\Phi(G)}2\text{i}\sin(-\pi\alpha(u))=\prod_{\alpha\in\Phi(G)}\left(-x^{-{\alpha}/{2}}\left(x^{\alpha}-1\right)\right),
    \end{equation}
    where $\Phi$ denotes the root system of $G$ and $x^{\alpha}=\text{e}^{2\pi\text{i}\alpha(u)}$. As we are considering the unitary gauge groups in this paper, we have
    \begin{align}
        Z_V=&\prod_{(i,j)}2\text{i}\sin\left(\pi u^{(a)}_i-\pi u^{(a)}_j\right)2\text{i}\sin\left(\pi u^{(a)}_j-\pi u^{(a)}_i\right)\nonumber\\
        =&\prod_{(i,j)}\left(x^{(a)}_ix^{(a)}_j\right)^{-1}\left(x^{(a)}_i-x^{(a)}_j\right)\left(x^{(a)}_j-x^{(a)}_i\right),\label{vect}
    \end{align}
    where the product is over all pairs of $(i,j)$ with $i\neq j$, and we have included the superscript $(a)$ to indicate the gauge node in the quiver.
    \item Chiral multiplet $\chi$ in a representation $\mathtt{R}$:
    \begin{equation}
        Z_{\chi}=\prod_{\rho\in\mathtt{R}}\frac{1}{\text{2i}\sin\left(\pi\rho(u)+\pi F(z)\right)}=\prod_{\rho\in\mathtt{R}}\frac{1}{x^{-{\rho}/{2}}y^{-{F}/{2}}\left(x^{\rho}y^F-1\right)}\;,
    \end{equation}
    where $F$ denotes the flavour charge. For the chirals connecting two unitary gauge nodes $a,b$ of ranks $N_a,N_b$, we have\footnote{The first line comes from the fact that the adjoint representation of U$(N)$ is reducible. In other words, there is a U$(1)$ part besides the SU$(N)$ roots.}
    \begin{align}
        Z_{\chi}
        &=\left(\frac{1}{\prod\limits_I2\text{i}\sin\left(\pi F_{\chi_{aa,I}}(\epsilon_k)\right)}\right)^{\delta_{ab}N_a}\nonumber\\
        &\times\prod_{i=1}^{N_a}\prod_{j=1}^{N_b}\frac{1}{\prod\limits_I2\text{i}\sin\left(\pi u^{(a)}_i-\pi u^{(b)}_j+\pi F_{\chi_{ba,I}}(\epsilon_k)\right)\prod\limits_I2\text{i}\sin\left(\pi u^{(b)}_j-\pi u^{(a)}_i+\pi F_{\chi_{ab,I}}(\epsilon_k)\right)}\nonumber\\
        &=\left(\frac{1}{\prod\limits_Iq_{\chi_{aa,I}}^{-\frac{1}{2}}\left(1-q_{\chi_{ba,I}}\right)}\right)^{\delta_{ab}N_a}\nonumber\\
        &\times\prod_{i=1}^{N_a}\prod_{j=1}^{N_b}\frac{1}{\prod\limits_I\left(x^{(a)}_ix^{(b)}_jq_{\chi_{ba,I}}\right)^{-{1}/{2}}\left(x^{(a)}_i-q_{\chi_{ba,I}}x^{(b)}_j\right)\prod\limits_I\left(x^{(a)}_ix^{(b)}_jq_{\chi_{ab,I}}\right)^{-{1}/{2}}\left(x^{(b)}_j-q_{\chi_{ab,I}}x^{(a)}_i\right)}\;,\label{chi}
    \end{align}
    where $\epsilon_I=-F_I\left(\epsilon_k\right)$. To avoid possible confusions, we shall always refer to $\epsilon_I$ and $F_I$ as weights and charges respectively so that they have the opposite signs\footnote{For the 2d $\mathcal{N}=(0,2)$ quivers that can be obtained from dimensional reduction of 4d theories, the weights can also be determined as follows. As the original periodic quiver (for 4d $\mathcal{N}=1$) on $T^2$ is uplifted to one on $T^3$, we have an extra cycle/direction parametrized by the new adjoints $\Phi_{aa}$. This would then cause the vertical shifts of the multiplets in the periodic quiver. Suppose that $\Phi_{aa}$ are shifted by 1. As a result, the superpotential would have shift $-1$ as the $J$- and $E$-terms should have opposite vertical shifts. We may then write the shift of $X_{ab}$ ($J_{ba}$, resp.~$E_{ab}$) as $s_{ab}$ ($-s_{ab}-1$, resp.~$s_{ab}+1$).

To determine $s_{ab}$, we can choose one perfect matching in the brane tiling of the 4d theory. As a perfect matching would pick out precisely one $X_{ab}$ in each monomial term of the superpotential, we can therefore choose these $X_{ab}$ in this perfect matching to have $s_{ab}=-1$ while keeping the other chirals unshifted. In the toric diagram, this adds a vertex above the one corresponding to the chosen perfect matching, uplifting the polygon into a polyhedron (Of course, choosing different perfect matchings would not change the geometry due to the SL$(3,\mathbb{Z})$ invariance of the toric diagram). Notice that the shifts of the Fermis $\Lambda_{ab}$ are given by $s_{ab}+1(\neq s_{ab})$, so the chirals and the Fermis would connect different (lattice) nodes in the periodic quiver.

Therefore, for these 2d theories, we can assign the weights of $\Phi_{aa}$, whose vertical shifts are always 1, to be $\epsilon_4$. The weights of $X_{ab}$ are then shifted by $s_{ab}\epsilon_4(=-\epsilon_4\text{ or }0)$ compared to their 4d $\mathcal{N}=1$ counterparts.}. We shall also use $w_i=\text{e}^{2\pi\text{i}v_i}$ for the fields connected to the framing node.
    \item Fermi multiplet $\Lambda$ in a representation $\mathtt{R}$:
    \begin{equation}
        Z_{\Lambda}=\prod_{\rho\in\mathtt{R}}2\text{i}\sin\left(-\pi\rho(u)-\pi F(z)\right)=\prod_{\rho\in\mathtt{R}}\left(-x^{-{\rho}/{2}}y^{-{F}/{2}}\left(x^{\rho}y^F-1\right)\right)\;.
    \end{equation}
    For the Fermis connecting two unitary gauge nodes $a,b$ of ranks $N_a,N_b$, we have
    \begin{align}
        Z_{\Lambda}=&\left(\prod\limits_I2\text{i}\sin\left(-\pi F_{\Lambda_I}(\epsilon_k)\right)\right)^{\delta_{ab}N_a}\prod_{i=1}^{N_a}\prod_{j=1}^{N_b}\prod\limits_I2\text{i}\sin\left(-\pi u^{t(\Lambda_I)}+\pi u^{s(\Lambda_I)}-\pi F_{I}(\epsilon_k)\right)\nonumber\\
        =&\left(\prod\limits_I\left(-q_I^{-{1}/{2}}\left(1-q_I\right)\right)\right)^{\delta_{ab}N_a}\prod_{i=1}^{N_a}\prod_{j=1}^{N_b}\prod\limits_I\left(-\left(x^{t(\Lambda_I)}x^{s(\Lambda_I)}q_I\right)^{-{1}/{2}}\left(x^{t(\Lambda_I)}-q_{I}x^{s(\Lambda_I)}\right)\right)\;,\label{fermi}
    \end{align}
    where $s(\Lambda),t(\Lambda)\in\{a,b\}$ based on the choice of the orientations of the edges\footnote{Recall that choosing an orientation for one edge would simultaneously determine the orientations for all the other edges. Different choices should give the same result due to the symmetry between $\Lambda$ and $\overline{\Lambda}$.}, and hence we have also omitted the subscripts $i,j$.
\end{itemize}

\paragraph{The space $\mathfrak{M}$} For gauge group $G$ of rank $N$ whose Lie algebra is $\mathfrak{g}$, denote the Cartan subalgebra as $\mathfrak{h}$ and the coroot lattice as $Q^{\vee}$. Then the space $\mathfrak{M}$ is defined as $\mathfrak{h}_{\mathbb{C}}/Q^{\vee}$.
From $Z_\text{1-loop}$, each multiplet gives rise to a hyperplane $H_i=\{Q_i(u)+\dots=0\}\subset\mathfrak{M}$ with covector $Q_i\in\mathfrak{h}^*$. For instance, a chiral leads to $\rho(u)+F(z)=0$ where $Q_i=\rho$.

Take $\mathfrak{M}_\text{sing}=\bigcup\limits_iH_i$. The set of isolated points where at least $N$ linearly independent hyperplanes meet is denoted as $\mathfrak{M}_\text{sing}^*$. Then $\bm{Q}(u^*)$ is the set of $Q_i$ meeting at $u^*\in\mathfrak{M}_\text{sing}^*$. In this paper, the singularities are always non-degenerate. In other words, the number of hyperplanes meeting at $u^*$ is always $N$ for any $u^*$.

The covector $\eta\in\mathfrak{h}^*$ picks out the allowed sets of hyperplanes in the JK residue. This is given by the positivity condition:
\begin{equation}
    \eta\in\text{Cone}(Q_{i_j}):=\left\{\sum_{j=1}^Na_jQ_{i_j}\,\Bigg|\,a_j\geq0\right\} \;.
\end{equation}
Here, we shall mainly focus on the choice $\eta=(1,1,\dots,1)$, whose admissible singularities as we will see give rise to the crystal structure corresponding to the cyclic chambers.

\paragraph{The Jeffrey-Kirwan Residue} The JK residue \cite{MR1318878} (see also \cite{Witten:1992xu}) is defined by
\begin{equation}
    \text{JK-Res}\, \frac{\text{d}Q_{i_1}(u)}{Q_{i_1}(u)}\wedge\dots\wedge\frac{\text{d}Q_{i_N}(u)}{Q_{i_N}(u)}:=\begin{cases}
        \text{sgn}(\text{det}(Q_{i_1},\dots,Q_{i_N}))\;,&\eta\in\text{Cone}(Q_{i_j})\;,\\
        0\;,&\text{otherwise} \;,
    \end{cases}
\end{equation}
which can be rewritten as
\begin{equation}
    \text{JK-Res}\, \frac{\text{d}u_1\wedge\dots\wedge\text{d}u_N}{Q_{i_1}(u)\dots Q_{i_N}(u)}=\begin{cases}
        \displaystyle\frac{1}{|\text{det}(Q_{i_1},\dots,Q_{i_N})|} \;, &\eta\in\text{Cone}(Q_{i_j}) \;,\\
        0\;,&\text{otherwise} \;,
    \end{cases}.
\end{equation}

There is an equivalent way to define the JK residue constructively as a sum of iterated residues \cite{szenes2003toric}. The key ingredient is a flag. For each $u^*$ with $(Q_{i_1},\dots,Q_{i_N})$, we consider a flag
\begin{equation}
\mathcal{F} = [ \mathcal{F}_0=\{0\} \subset \mathcal{F}_1\subset \dots \mathcal{F}_N ],\quad \textrm{dim}\, \mathcal{F}_j = j
\end{equation}
such that the vector space $\mathcal{F}_j$ at level $j$ is spanned by $\{Q_{i_1},\dots,Q_{i_j}\}$. Denote the set of flags as $\mathcal{FL}(\bm{Q}(u^*))$, and we choose the subset
\begin{equation}
    \mathcal{FL}^+(\bm{Q}(u^*))=\left\{\mathcal{F}\in\mathcal{FL}(\bm{Q}(u^*))\,\bigg|\,\eta\in\text{Cone}\left(\kappa^{\mathcal{F}}_1,\dots,\kappa^{\mathcal{F}}_N\right)\right\}\;,
\end{equation}
where
\begin{equation}
    \kappa^{\mathcal{F}}_j :=\sum_{Q_i\in\bm{Q}(u^*)\cap\mathcal{F}_j}Q_i \;.
\end{equation}
Introduce the sign factor $\nu(\mathcal{F})=\text{sgn}\left(\text{det}\left(\nu^{\mathcal{F}}_1,\dots,\nu^{\mathcal{F}}_N\right)\right)$. The JK residue can then be obtained by
\begin{equation}
\textrm{JK-Res}(\bm{Q}(u^*),\eta) = \sum_{\mathcal{F}} \nu(\mathcal{F}) \textrm{JK-Res}_{\mathcal{F}} \;.
\end{equation}
Here, the iterated residue $\text{JK-Res}_{\mathcal{F}}$ is defined as follows. Given an $N$-form $\omega=\omega_{1,\dots,N}\text{d}u_1\wedge\dots\wedge\text{d}u_N$, choose new coordinates
\begin{equation}
    \widetilde{u}_j=Q_{i_j}\cdot u,\quad j=1,\dots,N 
\end{equation}
such that $\omega=\widetilde{\omega}_{1,\dots,N}\text{d}\widetilde{u}_1\wedge\dots\wedge\text{d}\widetilde{u}_N$. Once this is given, the contribution to the JK residue from a flag is evaluated as
\begin{equation}
\textrm{JK-Res}_{\mathcal{F}}\, \omega =  \Res_{\widetilde{u}_r=\widetilde{u}_r^*} \dots \Res_{\widetilde{u}_1=\widetilde{u}_1^*} \widetilde{\omega}_{1,\dots,N}=J\left(\frac{\partial\widetilde{u}_i}{\partial u_j}\right)\Res_{u_r=u_r^*} \dots \Res_{u_1=u_1^*} \omega_{1,\dots,N} \;,
\end{equation}
where $J$ denotes the Jacobian.

\paragraph{Remark} In \cite{Kimura:2023bxy}, the integrand for the instanton partition function of the gauge origami system \cite{Nekrasov:2016ydq} was expressed using certain vertex operators. The OPEs of the vertex operators can be read off from the quivers. Using the above expression for $Z_{\text{1-loop}}$, it is straightforward to verify the statement therein. These vertex operators can then be used to define the so-called quiver $\mathcal{W}$-algebras \cite{Kimura:2015rgi}.

\subsection{Flags and JK Residues}\label{crystal}
Although the JK residue formula and the computation of the index contain more information, the collection of the intersection points $u^*$ of the hyperplanes can be viewed as some 4d configuration for the CY fourfold. Each site in the configuration corresponds to one of the coordinates in $u^*$. For later convenience, we may call such configurations ``crystals'' and their sites ``atoms''. The first definition of the JK residue above tells us about the ``final state'' of the crystal. On the other hand, the constructive definition of the JK residue tracks the growth/melting of the crystal (say, from rank $N-1$ to rank $N$) revealed by the flags. Of course, the equivalence of the two definitions means that the index only depends on the final shape of the crystal configuration. In the followings, we shall derive the melting rule for the 4d crystals.

When evaluating the JK residue using flags, it would be helpful to write the weight at level $N$ as
\begin{equation}
    Z_N\left(u_1,\dots,u_N\right)=Z_{N-1}\left(u_1,\dots,u_{N-1}\right)\Delta Z_{N-1,N}\left(u_1,\dots,u_N\right)\;,
\end{equation}
where we have suppressed the superscripts $(a)$ of $u_i$ for brevity. As the poles are always simple, the iterated residues have the decomposition
\begin{align}
    &\text{Res}_{u_N=u_N^*}\dots\text{Res}_{u_1=u_1^*}Z_N\left(u_1,\dots,u_N\right)\nonumber\\
    &=\text{Res}_{u_N=u_N^*}\dots\text{Res}_{u_1=u_1^*}\left(Z_{N-1}\left(u_1,\dots,u_{N-1}\right)\Delta Z_{N-1,N}\left(u_1,\dots,u_N\right)\right)\nonumber\\
    &=\left(\text{Res}_{u_{N-1}=u_{N-1}^*}\dots\text{Res}_{u_1=u_1^*}Z_{N-1}\left(u_1,\dots,u_{N-1}\right)\right)\left(\text{Res}_{u_N=u_N^*}\Delta Z_{N-1,N}\left(u_1^*,\dots,u_{N-1}^*,u_N\right)\right)\;,
    \label{iteratedres}
\end{align}
and the first bracket in the last line is the one obtained at level $N-1$.

As $Z_N$ can be obtained from $Z_{N-1}$ by multiplying the factor $\Delta Z_{N-1,N}$. Together with the contributions from the vector multiplet
\begin{equation}
    \prod\limits_{i=1}^{N_a-1}2\text{i}\sin\left(\pi u^{(a)}_{N_a}-\pi u^{(a)}_i\right)2\text{i}\sin\left(\pi u^{(a)}_i-\pi u^{(a)}_{N_a}\right)\;,
    \label{vect}
\end{equation}
the extra $\Delta Z_{N-1,N}$ is composed of the factors\footnote{Here, we simply take one chiral and one Fermi from the framing node to one (initial) gauge node. One can of course consider more general framings with multiple edges connecting the framing node and different gauge nodes, some of which could be related by wall crossing.}
\begin{equation}
    \left(\frac{2\text{i}\sin\left(-\pi v_2+\pi u^{(a)}_{N_a}\right)}{2\text{i}\sin\left(\pi u^{(a)}_{N_a}-\pi v_1\right)}\right)^{\delta_{a,0}}\;,
    \label{framing}
\end{equation}
with the initial node connected to the framing node labelled by 0, and\footnote{We have omitted the U$(1)$ part contributions from the possible adjoint loops as they do not have any poles in $u$.}
\begin{equation}
    \prod_{i=1}^{N_b}\left(\frac{\prod\limits_I2\text{i}\sin\left(-\pi u^{t(\Lambda_I)}+\pi u^{s(\Lambda_I)}-\pi F_{\Lambda_{I}}(\epsilon_k)\right)}{\prod\limits_I2\text{i}\sin\left(\pi u^{(a)}_{N_a}-\pi u^{(b)}_i+\pi F_{\chi_{ba,I}}(\epsilon_k)\right)\prod\limits_I2\text{i}\sin\left(\pi u^{(b)}_i-\pi u^{(a)}_{N_a}+\pi F_{\chi_{ab,I}}(\epsilon_k)\right)}\right)\label{chiralfermi}
\end{equation}
for all nodes $b$ connected to the node $a$. Here, we require the quiver theory to satisfy the followings:
\begin{itemize}
    \item The theory has a corresponding periodic quiver, which is a weight lattice.
    \item There is at most one edge (either chiral or Fermi) connecting any two lattice points (namely, the nodes in the periodic quiver).
\end{itemize}
In paritcular, this is true for quivers arised from toric CY fourfolds considered in the paper\footnote{One way to see this is to consider the brane brick model of the quiver. The lattice points correspond to the brane bricks, i.e., the convex polytopes, in this dual graph. Any two polytopes can share at most one face, which is the edge in the weight lattice/periodic quiver.}. This indicates that the flavour charge of the chiral $\chi_{ab,I}$ (or $\chi_{ba,I}$) would not only differ the one of a Fermi $\Lambda_{ab,I}$ by a multiple of $\epsilon$. In other words, $F_{\chi_{ab,I}}\neq F_{\Lambda_{ab,I}}=F_{\Lambda_{ab,I}}+n\epsilon$ (recall that eventually we would like to take $\epsilon=0$). Likewise, none of the adjoint chirals would have flavour charge being a multiple of $\epsilon$ as well, i.e., $F_{\text{adj}}\neq0=n\epsilon$. This means that we can take $\epsilon=0$ (or equivalently, $q=1$) before evaluating the residues (as opposed to the threefold cases discussed in Appendix \ref{3dcrystals} where the order is reversed). In fact, the Calabi-Yau condition should always come before the evaluation of the integral so that the poles could be correctly cancelled by the factors in the numerator. For example, for the $\mathbb{C}^4$ case whose periodic quiver was given in \eqref{periodicquiverex}, at rank 4, there is one contribution from
\begin{equation}
    (u_1,u_2,u_3,u_4)=(v_1,u_1+\epsilon_3,u_1+\epsilon_4,u_2+\epsilon_4)=(v_1,u_1+\epsilon_3,u_1+\epsilon_4,u_3+\epsilon_3)\;.\label{orderex}
\end{equation}
This can be depicted as
\begin{equation}
    \includegraphics[width=4cm]{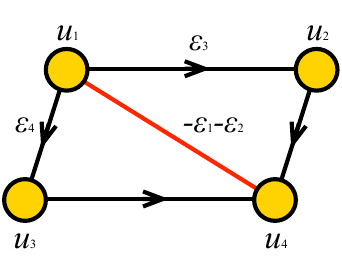}.
\end{equation}
For $u_4$, the corresponding pole has order 2, which comes from the contributions $u_4-u_2-\epsilon_3$ and $u_4-u_3-\epsilon_4$ (that are both equal to $u_4-v_1-\epsilon_3-\epsilon_4$) due to the chirals\footnote{Here, we are using the rational version as it is the most straightforward one to see how the pole structure is related to the periodic quiver.}. However, this is actually a simple pole since one such factor is cancelled by $u_1-u_4-\epsilon_1-\epsilon_2$ in the numerator due to the Fermi. This is because we need to take the Calabi-Yau condition $\epsilon_1+\epsilon_2+\epsilon_3+\epsilon_4=0$ first so that $u_1-u_4-\epsilon_1-\epsilon_2=-(u_4-u_1-\epsilon_3-\epsilon_4)$.

Notice that not all the flags have non-vanishing contributions.
In order to characterize the flags with non-vanishing contributions,
we first discuss the construction of the 4d crystals from the periodic quivers.

For the case of the Calabi-Yau threefold,
the process of obtaining the 3d crystal is as follows \cite{Ooguri:2008yb}:
\begin{enumerate}
    \item We start with a quiver diagram on the two-torus, and consider the periodic quiver in the universal cover, i.e., $\mathbb{R}^2$.
    \item Choose a vertex of the quiver diagram, and consider a set of paths on the quiver diagram.
    \item We can add an extra direction by considering the R-charges.
\end{enumerate}

\noindent
We follow the same strategy as in our case at hand:
\begin{enumerate}
    \item We start with a quiver diagram on the three-torus, and consider the periodic quiver in the universal cover, i.e., $\mathbb{R}^3$.
    \item Choose a vertex of the quiver diagram, and consider a set of (chiral) paths on the quiver diagram.
    \item We can uplift the 3d periodic quiver with the direction coming from the R-charges. Each path in the quiver diagram (starting from the initial node $a_0$ and ending at some node $b$) can be written as $p_{a_0,b}(n)=v_{a_0,b}\omega^n$. Here, $v_{a_0,b}$ is the shortest path from $a_0$ to $b$, and $\omega$ corresponds to some closed loop. This loop can be viewed as composed of the chirals only, or one can equivalently take some ``shortcut'' such that it consists of both chirals and Fermis due to the $J$-/$E$-terms. Following this path $p_{a_0,b}(n)$, we put an atom $b$ at depth $n$ in the uplift direction below the atom of colour $b$ with the path $p_{a_0,b}(0)=v_{a_0,b}$.
\end{enumerate}
We notice that combinatorially this is exactly the structure discussed in the very nice paper \cite{Franco:2023tly}. Readers are referred to \cite{Franco:2023tly} for more details on the 4d crystals and the connections to the brane brick models. Here, our interest is to derive the crystals from the perspective of the BPS states counting.

\subsection{Crystals from JK Residues}\label{crystalproof}

As the JK residues over the flags may or may not be zero, the crystals cannot grow arbitrarily. Now, let us derive the melting rule of the crystals. While this was done previously for the particular examples of the solid partitions in the $\mathbb{C}^4$ \cite{Nekrasov:2018xsb}, we will generalize the discussion to an arbitrary toric Calabi-Yau four-fold.

For brevity, let us denote the basis vector $(0,\dots,0,1,0,\dots,0)\in\mathbb{R}^N$ with the only non-zero element at the $i^\text{th}$ entry as $\bm{e}_i$. Then $\eta=\sum\limits_{i=1}^N\bm{e}_i$ at level $N$. For an admissible set $\sigma$ such that $\eta\in\text{Cone}(\sigma)$, we can see a tree structure  as follows. Given that $\sigma$ must contain at least one $\bm{e}_j$, the vector $\bm{e}_k-\bm{e}_j$ is allowed in $\sigma$ while $\bm{e}_j-\bm{e}_k$ is not. Then if $\bm{e}_j$ and $\bm{e}_k-\bm{e}_j$ are in $\sigma$, the vector $\bm{e}_l-\bm{e}_k$ is allowed while $\bm{e}_k-\bm{e}_l$ is not etc.

At a generic level, there would still be more admissible $\sigma$ than the atoms in the crystal due to the cancellations of the factors in the numerator and the denominator in $Z_\text{1-loop}$. 

Our proof proceeds by induction with respect to the rank of the gauge group.
Suppose that the pole structures are consistent with the crystal configurations at level $N-1$. Then we can focus on the part $\Delta Z_{N-1,N}$ and always assume that $\sigma=\sigma^{(N-1)}\sqcup\{\bm{e}_N-\bm{e}_j\}$ for some $j<N$ where $\sigma^{(N-1)}$ only involves $\bm{e}_{i<N}$.

In $\Delta Z_{N-1,N}$, the possible poles would be in one of the following scenarios:
\begin{itemize}[leftmargin=*]
    \item If the pole for $u^{(a)}_N$ corresponds to an atom already appeared in the molten crystal at size $N-1$, then $u^{(a)}_N=u^{(b)}_j-F_{\chi_{ba,I}}(\epsilon_k)=u^{(a)}_{j'}$ for some $j'<N$ as depicted in Fig.~\ref{scenario1}.
    \begin{figure}[ht]
        \centering
        \includegraphics[width=4cm]{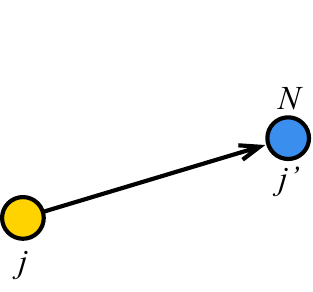}
        \caption{Suppose the atom labelled $N$ is already existing in the crystal. The pole $u^{(a)}_N=u^{(b)}_j-F_{\chi_{ba,I}}$ corresponds to the arrow from $j$ to $N$. However, this atom already has some label $j'$ as it has been added to the crystal.}\label{scenario1}
    \end{figure}
    The pole containing $u^{(a)}_N-u^{(a)}_{j'}$ would then be cancelled by the same factor in the numerator coming from the contribution of the vector multiplet as given in \eqref{vect}.

    \item If the pole for $u^{(a)}_N$ corresponds to an addable atom for the molten crystal, then $u^{(a)}_N=u^{(b)}_j-F_{\chi_{ba,I}}(\epsilon_k)$ gives a pole for some $j<N$. It could be possible that there are multiple such factors, namely $u^{(a)}_N=u^{(b_1)}_{j_1}-F_{\chi_{b_1a,I}}(\epsilon_k)=u^{(b_2)}_{j_2}-F_{\chi_{b_2a,J}}(\epsilon_k)=\dots$. There would exist some $j'$ such that
    \begin{equation}
        u^{(a)}_N=u^{(b_1)}_{j_1}-F_{\chi_{b_1a,I}}(\epsilon_k)=u^{(c)}_{j'}-F_{\chi_{cd_1,K}}(\epsilon_k)-\dots-F_{\chi_{d_nb_1,L}}(\epsilon_k)-F_{\chi_{b_1a,I}}(\epsilon_k)=u^{(c)}_{j'}-F_{\Lambda_{ab,I}}(\epsilon_k)\;,
        \label{polecancellation}
    \end{equation}
    where the last equality follows from the $J$-/$E$-term plaquette in the quiver. As a result, one of the two factors from $u^{(a)}_N=u^{(b_1)}_{j_1}-F_{\chi_{b_1a,I}}=u^{(b_2)}_{j_2}-F_{\chi_{b_2a,I}}$ is killed by the corresponding contribution from the Fermi multiplet \eqref{fermi} in the numerator. After cancelling one such factor, the surviving one with the next factor, say from $u^{(b_3)}_{j_3}$, would form another pair, and this procedure can be repeated pairwise. Eventually, there would be a simple pole left. In other words, the number of chirals connecting the existing atoms and an addable atom is always one more than the number of Fermis connecting the existing atoms and this addable atom in the crystal\footnote{Of course, the initial atom(s) would be special as we keep the weights of the edges connecting the framing node generic so that the corresponding factors in the numerator and the denominator do not cancel each other.}. Schematically, this is shown in Fig.~\ref{scenario2}.
    \begin{figure}
        \centering
        \includegraphics[width=10cm]{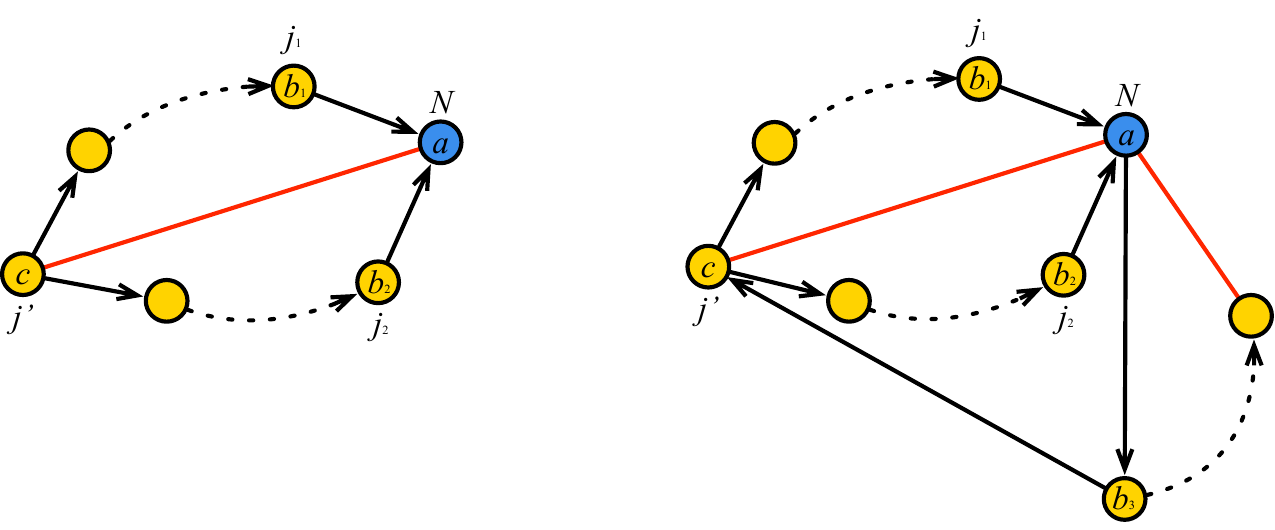}
        \caption{The atom to be added at level $N$ is shown in blue. Left: There are two same factors in the denominator coming from the chirals, but one of them gets cancelled by the contribution from the Fermi. This is in fact a pair of plaquettes from the $J$- or $E$-term. Right: There are multiple contributions from the chirals (either pointing forwards or pointing backwards) to the same pole. This would be cancelled to a simple pole by the factors in the numerator from the Fermis.}\label{scenario2}
    \end{figure}

    \item If the pole for $u^{(a)}_N$ corresponds to a position that does not belong to the size $N$ molten crystal (obtained from the size $N-1$ crystal), then the cancellation of this factor is the same as given in \eqref{polecancellation}. This is diagrammatically shown in Fig.~\ref{scenario3}.
    \begin{figure}[ht]
        \centering
        \includegraphics[width=5cm]{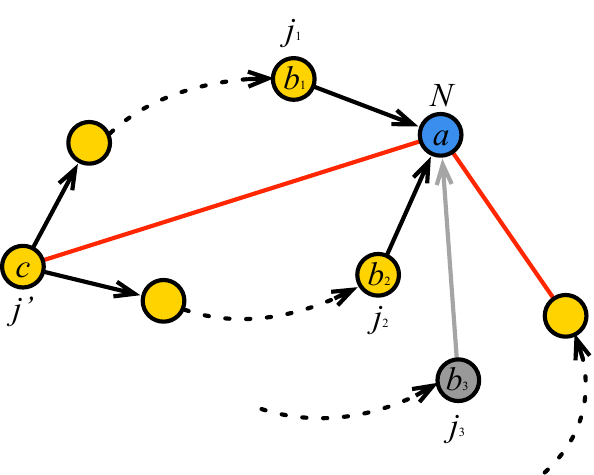}
        \caption{The atom and the arrow in grey are not present in the crystal. Therefore, the blue atom cannot be added to the crystal as there is no corresponding pole.}\label{scenario3}
    \end{figure}
\end{itemize}
Therefore, the 4d configuration at level $N$ would be a crystal of size $N$. By induction, the fixed points are in one-to-one correspondence with the crystal configurations satisfying the melting rule: an atom $\mathfrak{a}$ is in the molten crystal $\mathfrak{C}$ whenever there exists a chiral $I$ such that $I\cdot\mathfrak{a}\in\mathfrak{C}$. In other words, $I\cdot\mathfrak{a}\notin\mathfrak{C}$ whenever $\mathfrak{a}\notin\mathfrak{C}$. If one considers the Jacobi algebra $\mathbb{C}Q/\langle J\text{-},E\text{-terms}\rangle$, the melting rule says that the complement of $\mathfrak{C}$ is its ideal.

\paragraph{Example} Let us illustrate the above three scenarios with an example. For $\mathbb{C}^4$ whose toric diagram and quiver can be found in \eqref{C4} below, the 1-loop determinant reads
\begin{align}
    Z_\text{1-loop}=&\left(\frac{\prod\limits_{1\leq k<l\leq3}(-\epsilon_k-\epsilon_l)}{\prod\limits_{k=1}^4(-\epsilon_k)}\right)^N\left(\prod_{i=1}^N\frac{(u_i-v_2)}{(u_i-v_1)}\text{d}u_i\right)\nonumber\\
    &\left(\prod_{i\neq j}^N\frac{(u_j-u_i)\prod\limits_{1\leq k<l\leq3}(u_i-u_j-\epsilon_k-\epsilon_l)}{\prod\limits_{k=1}^4(u_j-u_i-\epsilon_k)}\right)\;,
\end{align}
where we have used the rational version for simplicity. Let us give one example for each of the scenario:
\begin{itemize}
    \item Suppose we have a single atom with position $u_1=v_1$ in the crystal, and we would like to add the second one to it. We have
    \begin{align}
    Z_\text{1-loop}=&\left(\frac{\prod\limits_{1\leq k<l\leq3}(-\epsilon_k-\epsilon_l)}{\prod\limits_{k=1}^4(-\epsilon_k)}\right)^2\left(\frac{(u_1-v_2)}{(u_1-v_1)}\frac{(u_2-v_2)}{(u_2-v_1)}\text{d}u_1\text{d}u_2\right)\nonumber\\
    &\left(\prod_{i\neq j}^2\frac{(u_j-u_i)\prod\limits_{1\leq k<l\leq3}(u_i-u_j-\epsilon_k-\epsilon_l)}{\prod\limits_{k=1}^4(u_j-u_i-\epsilon_k)}\right)\;.
\end{align}
Then the contribution from $u_2=v_1$ is zero since the pole $u_2-v_1$ is cancelled by the factor $u_2-u_1=u_2-v_1$ in the numerator.

\item Consider the figure in \eqref{orderex} which is reproduced here:
\begin{equation}
    \includegraphics[width=4cm]{figures/orderex.pdf}.
\end{equation}
Suppose we have the configuration with atoms labelled by $(u_1,u_2,u_3)$. To add the atom labelled by $u_4$, we have two poles
\begin{equation}
    u_4-u_2-\epsilon_3=u_4-u_1-\epsilon_3-\epsilon_4\text{ and }u_4-u_3-\epsilon_4=u_4-u_1-\epsilon_4-\epsilon_3.
\end{equation}
However, there is a factor
\begin{equation}
    u_4-u_1-(-\epsilon_1-\epsilon_2)=u_4-u_1-\epsilon_3-\epsilon_4
\end{equation}
from the Fermi multiplet in the numerator. Therefore, we would have a simple pole left as expected.

\item In this figure, the configuration with only $(u_1,u_2,u_4)$ is not allowed since the corresponding pole $u_4-u_2-\epsilon_4$ would be cancelled by $u_4-u_1-(-\epsilon_1-\epsilon_2)=$ in the numerator from the Fermi multiplet. Likewise, the configuration $(u_1,u_3,u_4)$ is also not allowed. Only when both $u_2=u_1+\epsilon_3$ and $u_3=u_1+\epsilon_4$ are present can we add $u_4$ in the above figure.
\end{itemize}

\subsection{Weights for Crystals}\label{weights}
For a toric CY$_3$, each crystal corresponds to a BPS state in the index up to a sign\footnote{We also show this using the JK residue formula in Appendix \ref{3dcrystals}.}. However, as mentioned before, the 4d crystals, albeit having a one-to-one correspondence with the fixed points, do not encode the full information of the BPS spectrum.

In general, it is not easy to write down the full BPS partition function in a closed form. Nevertheless, using the constructive definition of the JK residue, we can get the recursive formula for the BPS indices. Recall that the BPS partition function reads\footnote{Here, we use the formal variable $p_a$ representing the colours of the crystal. There should be non-trivial maps from $p_a$ to the variables associated to the D-branes.}
\begin{equation}
    \mathcal{Z}_{\text{BPS}}(p_0,\dots,p_{|Q_0|-1})=\sum_{n_0+\dots+n_{|Q_0|-1}=N}\mathcal{I}_n(q_k)p_0^{n_0}\dots p_{|Q_0|-1}^{n_{|Q_0|-1}}\;,
\end{equation}
where $\mathcal{I}_N(q_k)$ is the index at level $N$ which takes all the crystal configurations of size $N$ into account, but with non-trivial weights depending on $q_k$. In other words, we have
\begin{equation}
    \mathcal{I}_N(q_k)=\sum_{|\mathfrak{C}|=N}\mathtt{Z}_{\mathfrak{C}}(q_k)\;,
\end{equation}
Recall that a crystal $\mathfrak{C}_N$ of size $N$ can be obtained from some crystal $\mathfrak{C}_{N-1}$ of size $N-1$. For brevity, let us abbreviate $\mathtt{Z}_{\mathfrak{C}_N}$ as $\mathtt{Z}_N$. Then
\begin{equation}
    \mathtt{Z}_{N}(q_k)=\mathtt{Z}_{N-1}(q_k)\Delta\mathtt{Z}_{N-1,N}(q_k)\;,
\end{equation}
where
\begin{align}
    &\mathtt{Z}_{N-1}(q_k)=\text{Res}_{u_{N-1}=u_{N-1}^*}\dots\text{Res}_{u_1=u_1^*}Z_{N-1}\left(u_1,\dots,u_{N-1}\right)\;,\nonumber\\
    &\Delta\mathtt{Z}_{N-1,N}(q_k)=\text{Res}_{u_N=u_N^*}\Delta Z_{N-1,N}\left(u_1^*,\dots,u_{N-1}^*,u_N\right)\;.
\end{align}
In particular, $u_i^*$ is actually a function of the fugacities $q_k$. This is precisely \eqref{iteratedres}. More concretely, for unitary gauge groups, the iterative factor $\Delta\mathtt{Z}_{N-1,N}(q_k)$ is given by
\begin{equation}
    \Delta\mathtt{Z}_{N-1,N}(q_k)=(\eqref{vect}\times\eqref{framing}\times\eqref{chiralfermi})'|_{u=u^*(q_k)}\;,
\end{equation}
where the apostrophe indicates that we remove the factor of the corresponding pole in the denominator (which always comes from the chirals).

\subsection{Wall Crossing}\label{wallcrossing}
For CY threefolds, as pointed out in \cite{Aganagic:2010qr}, the chambers under ``the wall crossing of the second kind'' \cite{Kontsevich:2008fj} are essentially related by Seiberg dualities. The cyclic chambers achieved this way still have the crystal structures, but with different shapes\footnote{Of course, the cyclic chambers are not necessarily obtained by mutations. See for example the dP$_0$ case in \cite[\S5.2]{Galakhov:2021xum}.}. Here, we can study the similar phenomena under mutations\footnote{Mathematically, quiver mutations are mainly defined for quivers without self-loops and 2-cycles (despite some extensions in some mathematical literature). Moreover, these quivers only have oriented arrows (namely no Fermis). In this paper, we simply use mutation to refer to the IR equivalence manipulation on the quivers.} on the node(s) connected to the framing node.

\begin{figure}[ht]
    \centering
    \includegraphics[width=16cm]{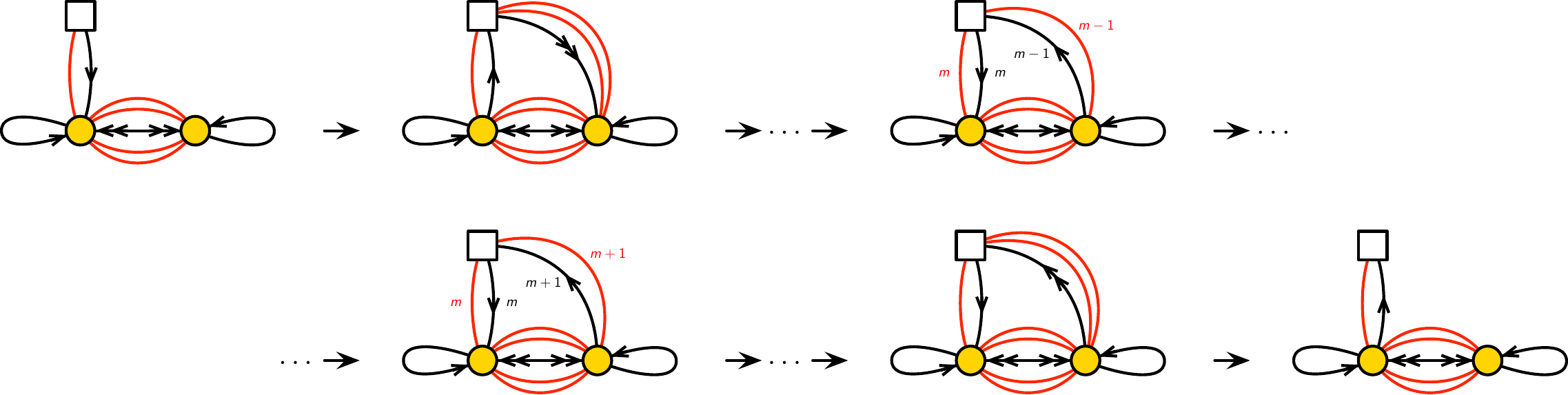}
    \caption{The wall crossing of the conifold$\times\mathbb{C}$ quiver. Here, we are only showing the chambers that descend from the cyclic chambers of the 4d $\mathcal{N}=1$ quiver. The labels indicate the multiplicities of the edges. These were also recently studied in \cite{Cao:2020huo}.}
    \label{wallcrossingex}
\end{figure}

In general, for an $\mathcal{N}=(0,2)$ quiver, we can mutate the quiver in line with the triality following the rules in \S\ref{review_1}. This would lead us to different cyclic chambers as the framing changes. However, for a 2d quiver with a 4d parent theory, we can also first take the Seiberg duality for the 4d quiver and then dimensionally reduce it to 2d. An example can be found in Figure \ref{wallcrossingex}.
We expect this to be different from the direct manipulation on the 2d quiver itself. In other words, the diagram
\begin{equation}
\begin{matrix}
    \includegraphics[width=8cm]{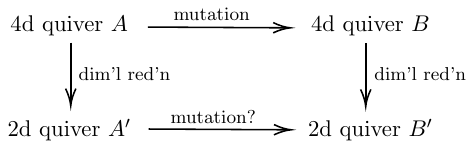}
\end{matrix}
\end{equation}
does not commute. While the general rules of triality/mutation for quivers with supersymmetry enhanced to $\mathcal{N}=(2,2)$ is still not clear, there seems to be richer chamber structures for toric CY fourfolds (even just for those admitting crystal descriptions).

Again, a closed expression for the wall crossing formula is not easy to obtain. Nevertheless, we can still use the JK residue formula to compute the index after the wall crossing. The new crystals can then be read off from the flags. The crystals can also be obtained directly from the quiver after the wall crossing, where the nodes/atoms connected to the framing node by the chirals correspond to the starters, pausers or stoppers in the language of \cite{Galakhov:2021xum}. The positions of these atoms are determined by the weights of the chirals. We shall consider more examples in \S\ref{ex}.

Moreover, one may also consider the covectors $\eta$ other than $(1,\dots,1)$. This would give rise to different sets of allowed cones. In particular, the resulting chambers may not have the combinatorial crystal descriptions satisfying the melting rule. Nevertheless, the BPS index can still be computed using the JK residues.

\section{Elliptic and Rational Generalizations}\label{ellipticandrational}
We have seen that the BPS index of a 1d quantum mechanics can be obtained from the 2d $\mathcal{N}=(0,2)$ elliptic genus under dimensional reduction. It is therefore natural to consider more general possibilities, by considering the 2d elliptic genera themselves (``elliptic version''), as opposed to 
the 1d indices (``trigonometric/K-theoretic  version''). We can also consider further dimensional reduction on a circle to 
0d, to discuss ``rational/cohomological version''.
Let us briefly discuss these in this section.

\subsection{The Elliptic Invariants}\label{elliptic}
 Instead of supersymmetric indices of 1d quantum mechanics, we can compute the 2d elliptic genus of the theory, by using the expressions for the 1-loop determinant reviewed in Appendix \ref{ellg}. We then obtain the elliptic version of the generalized DT invariants for the 2d $\mathcal{N}=(0,2)$ theories. By taking the trigonometric limit of the elliptic genus, one recovers the K-theoretic invariants discussed above. Note that some particular examples are discussed in the literature:
 for a special example of $\mathbb{C}^3$, the $\mathcal{N}=(2,2)$ elliptic invariants were computed in \cite{Benini:2018hjy} and dubbed ``elliptic DT invariants'',  and the cases with abelian gauge groups were considered in \cite{Franco:2017cjj}. Our discussion, however, applies to a general toric Calabi-Yau threefolds and fourfolds, and to a general non-Abelian theory.

There is one subtlety in the discussion of the 2d elliptic genus: unlike the 1d quiver quantum mechanics and the 0d quiver matrix model (to be mentioned below), the 2d theories are subject to anomaly constraints. This is similar to the case studied in \cite{Benini:2018hjy}.

Due to the gauge R-symmetry anomaly, the integrand of the elliptic genus is only doubly quasi-periodic under $u_i\rightarrow u_i+a+b\tau$ with $a,b\in\mathbb{Z}$ and any $u_i$ for generic $v_{1,2}$. Often, there would be an extra phase $\text{e}^{-2\pi\text{i}b(v_1-v_2)}$ under the transformation. Therefore, we need
\begin{equation}
    v_1-v_2\in\mathbb{Z} \;.
\end{equation}
Here, we are just considering the case when there is one chiral-Fermi pair from the framing node to only one of the gauge nodes\footnote{Of course, the flavour symmetry is U$(1)$ (or U$(1|1)$) in the context of (generalized) DT theory. For general SU$(M)$ (or SU$(M|M)$) flavour symmetry which can be thought of as multiple D8-branes, the condition would become $M(v_1-v_2)\in\mathbb{Z}$.}. One may also consider more general framings which could lead to different conditions depending on the cases. We will analyze some examples in \S\ref{ex}.

It is worth noting that under the shift $\epsilon_k\rightarrow\epsilon_k+1$ or $v_k\rightarrow v_k+1$ for any $\epsilon_k$ or $v_k$, the corresponding fugacity is invariant. However, the elliptic genus may not be so due to the presence of the Jacobi theta functions. Physically, this is caused by the 't Hooft anomaly, and $Z_{T^2}\rightarrow\pm Z_{T^2}$.

\subsection{The Rational Limit}\label{rational}
We may dimensionally reduce the 1d $\mathcal{N}=2$ quantum mechanics to a 0d quiver matrix model. This can be achieved by taking the rational limit $\beta\rightarrow0$ in the fugacities
\begin{equation}
    x_i=\text{e}^{2\pi\text{i}\beta u_i}\;,\quad w_i=\text{e}^{2\pi\text{i}\beta v_i}\;,\quad q_k=\text{e}^{2\pi\text{i}\beta\epsilon_k}\;,
\end{equation}
where we have taken the redefinition of the variables to make the scaling more explicit. In the index formula, this replaces the functions of form $\sin(u)$ with $u$. As a result, we have a partition function for the matrix model (the 0d theory).

Now that we have mentioned the 0d theories, it is tempting to consider those arising from a toric CY$_5$ and wonder if they would also admit some combinatorial structure described by 5d crystals for their partition functions (possibly with non-trivial weights). We run into a problem for a general toric CY$_5$---the resulting 0d theory has minimally $\mathcal{N}=1$ supersymmetry and we do not have a supersymmetric partition function via localization for 0d $\mathcal{N}=1$ theories. Nevertheless, for the 0d theories obtained from dimensional reductions of the 1d $\mathcal{N}=2$ theories as discussed above, we have more ($\mathcal{N}\ge 2$) supersymmetries. In other words, the formula of the partition function from the BPS index provides a partial combinatorial structure for the theories associated to $\mathbb{C}\times\text{CY}_4$. This is characterized by the same 4d crystals from their 1d parent theories (with non-trivial weights), where one parameter, say $\epsilon_5$, is turned off.

\section{Examples}\label{ex}
Let us now consider some examples as illustrations of our discussions above. Some features of some examples were also studied previously in \cite{Nekrasov:2017cih,Nekrasov:2018xsb,Bonelli:2020gku,Szabo:2023ixw,Kimura:2023bxy}.

\subsection{Solid Partitions: \texorpdfstring{$\mathbb{C}^4$}{C4}}\label{C4}
The simplest example would be the $\mathbb{C}^4$ case whose toric diagram and quiver are
\begin{equation}
\begin{matrix}
    \includegraphics[trim=0 30 0 -10,width=8cm]{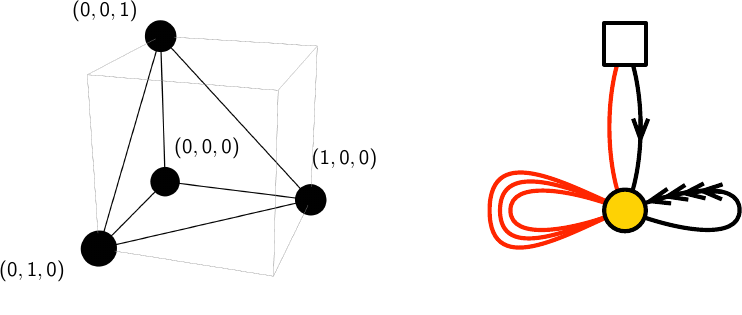}.\label{C4}
\end{matrix}
\end{equation}
It is straightforward to see that the four adjoint chirals\footnote{This can also be determined by the dimensional reduction for the $\mathbb{C}^3$ case where the three chirals have weights $\epsilon_1$, $\epsilon_2$ and $\epsilon_3'=-\epsilon_1-\epsilon_2$. After taking the vertical shift, the four adjoint chirals in the $\mathbb{C}^4$ quiver would have weights $\epsilon_1$, $\epsilon_2$, $\epsilon_3'-\epsilon_4$ and $\epsilon_4$. The identification $\epsilon_3=\epsilon_3'-\epsilon_4$ yields the weight assignment.} have weights $\epsilon_{1,2,3,4}$, and we shall pick the choice such that the weights of the three adjoint Fermis are $-\epsilon_k-\epsilon_l$ with $1\leq k<l\leq3$. We shall take the chiral (resp.~Fermi) connected to the framing node to have weight $v_1$ (resp.~$-v_2$)\footnote{When writing the 1-loop determinant, we shall always make the choice that the Fermis connected to the framing node are opposite to the accompanied chirals.}.

At rank $N$, the integrand is then
\begin{align}
    Z_\text{1-loop}
    &= q_4^{N^2}\left(\frac{\prod\limits_{1\leq k<l\leq3}(1-q_kq_l)}{\prod\limits_{k=1}^4(1-q_k)}\sqrt{\frac{w_1}{w_2}}\right)^N\left(\prod_{i=1}^N\frac{(x_i-w_2)}{(x_i-w_1)}\frac{\text{d}x_i}{x_i}\right)\nonumber\\
    &\left(\prod_{i\neq j}^N\frac{(x_j-x_i)\prod\limits_{1\leq k<l\leq3}(x_i-x_jq_kq_l)}{\prod\limits_{k=1}^4(x_j-x_iq_k)}\right).
\end{align}
It would also be convenient to write
\begin{equation}
    \mu :=w_1/w_2.
\end{equation}
Let us list some indices at low ranks as an illustration\footnote{Recall that we have $q\equiv q_1q_2q_3q_4=1$.}:
\begin{itemize}
    \item Level 1:
    \begin{itemize}
        \item crystal labelled by $v_1$:
        \begin{equation}
            \mathcal{I}_1=\frac{(1-q_1q_2)(1-q_1q_3)(1-q_2q_3)q_4\left(\sqrt{\mu}-1/\sqrt{\mu}\right)}{(1-q_1)(1-q_2)(1-q_3)(1-q_4)}\;.
        \end{equation}
    \end{itemize}
    \item Level 2:
    \begin{itemize}
        \item crystal labelled by $v_1,v_1+\epsilon_i$ ($i=1,2,3,4$):
        \begin{align}
            \mathcal{I}_{2,i}=&-\mathcal{I}_1\frac{(1-q_1q_2)(1-q_1q_3)(1-q_2q_3)q_4^3}{(1-q_1)(1-q_2)(1-q_3)(1-q_4)}(q_i\sqrt{\mu}-1/\sqrt{\mu})\nonumber\\
            &\frac{(q_j-1)(q_k-1)(q_l-1)(q_i^2q_j-1)(q_i^2q_k-1)(q_i^2q_l-1)}{(q_i+1)(q_i-q_j)(q_i-q_k)(q_i-q_l)(q_iq_j-1)(q_iq_k-1)(q_iq_l-1)}\;,
        \end{align}
        where $i,j,k,l\in\{1,2,3,4\}$ are distinct values (notice that $j,k,l$ are symmetric).
    \end{itemize}
    The index is
        \begin{equation}
            \mathcal{I}_2=\sum_{i=1}^4\mathcal{I}_{2,i}\;.
        \end{equation}
    \item Level 3:
    \begin{itemize}
        \item crystal labelled by $v_1,v_1+\epsilon_i+2\epsilon_i$ ($i=1,2,3,4$):
        \begin{align}
            \mathcal{I}_{3,(i,i)}=&-\mathcal{I}_{2,i}\frac{(1-q_1q_2)(1-q_1q_3)(1-q_2q_3)q_4^5}{(1-q_1)(1-q_2)(1-q_3)(1-q_4)}(q_i^2\sqrt{\mu}-1/\sqrt{\mu})\nonumber\\
            &\frac{(q_j-1)(q_k-1)(q_l-1)(q_i^3q_j-1)(q_i^3q_k-1)(q_i^3q_l-1)}{(q_i^2+q_i+1)(q_i^2-q_j)(q_i^2-q_k)(q_i^2-q_l)(q_iq_j-1)(q_iq_k-1)(q_iq_l-1)}\;,
        \end{align}
        where $i,j,k,l\in\{1,2,3,4\}$ are distinct values (notice that $j,k,l$ are symmetric);
        \item crystal labelled by $v_1,v_1+\epsilon_i+\epsilon_j$ ($i,j=1,2,3,4$, $i\neq j$):
        \begin{align}
            \mathcal{I}_{3,(i,j)}=&-\mathcal{I}_{2,i}\frac{(1-q_1q_2)(1-q_1q_3)(1-q_2q_3)q_4^5}{(1-q_1)(1-q_2)(1-q_3)(1-q_4)}(q_j\sqrt{\mu}-1/\sqrt{\mu})\nonumber\\
            &\frac{(q_i+1)(q_j-q_i)(q_i-1)(q_k-1)(q_l-1)(q_iq_j^2-1)(q_j^2q_k-q_i)(q_j^2q_l-q_i)}{(q_j^2-q_i)(q_j-q_i^2)(q_iq_j-1)(q_jq_k-q_i)(q_jq_l-q_i)(q_j-q_k)(q_j-q_l)}\;,
        \end{align}
        where $i,j,k,l\in\{1,2,3,4\}$ are distinct values (notice that $k,l$ are symmetric and $\mathcal{I}_{3,(i,j)}=\mathcal{I}_{3,(j,i)}$).
    \end{itemize}
    The index is
    \begin{equation}
        \mathcal{I}_3=\sum_{1\leq i\leq j\leq4}\mathcal{I}_{3,(i,j)}\;,
    \end{equation}
\end{itemize}

In general, we may also try to write down the generating function of the BPS indices\footnote{Often in literature, the subscripts of $p_a$ start from 0. Here, we save the label 0 for the framing node, and hence $a=1,\dots,|Q_0|$ for $p_a$.}
\begin{equation}
    Z_\text{BPS}=\sum_{N\in\mathbb{Z}_{\geq0}}\sum_{N_1+\dots+N_{|Q_0|}=N}\mathcal{I}_{N_1,\dots,N_{|Q_0|}}(q_k)p_1^{N_1}\dots p_{|Q_0|}^{N_{|Q_0|}}\;,
\end{equation}
where $p_a$ is the formal variable for each colour of the atoms in the crystal. It would also be convenient to introduce $p=(-1)^{|Q_0|}p_1p_2\dots p_{|Q_0|}$ (notice the sign).

For the $\mathbb{C}^4$ case, it was conjectured in \cite{Nekrasov:2017cih,Nekrasov:2018xsb} that the BPS partition function reads
\begin{align}
    &Z_\text{BPS}=\text{PE}[\mathcal{F}](p,\mu,\{q_k\})\;,\\
    &\mathcal{F}(p,\mu,\{q_k\})=\frac{[q_1q_2][q_1q_3][q_2q_3][\mu]}{[q_1][q_2][q_3][q_4][p\sqrt{\mu}][p/\sqrt{\mu}]}\;,
\end{align}
where $[X]:=X^{1/2}-X^{-1/2}$ and PE$[f](x_1,\dots,x_n)$ is the plethystic exponential of a function $f$ in variables $x_1,\dots,x_n$:
\begin{equation}
    \text{PE}[f](x_1,\dots,x_n):=\exp\left(\sum_{m=1}^{\infty}\frac{1}{m}f\left(x_1^m,\dots,x_n^m\right)\right)\;.
\end{equation}

\paragraph{Rational limit} In the rational limit, the integrand is
\begin{align}
    Z_\text{1-loop}=&\left(\frac{\prod\limits_{1\leq k<l\leq3}(-\epsilon_k-\epsilon_l)}{\prod\limits_{k=1}^4(-\epsilon_k)}\right)^N\left(\prod_{i=1}^N\frac{(u_i-v_2)}{(u_i-v_1)}\text{d}u_i\right)\nonumber\\
    &\left(\prod_{i\neq j}^N\frac{(u_j-u_i)\prod\limits_{1\leq k<l\leq3}(u_i-u_j-\epsilon_k-\epsilon_l)}{\prod\limits_{k=1}^4(u_j-u_i-\epsilon_k)}\right)\;.
\end{align}
The above partition functions then becomes
\begin{align}
    &Z_\text{mat}=\text{PE}[\mathcal{F}](p,v_1-v_2,\{\epsilon_k\})=M(p)^{\frac{(v_1-v_2)(\epsilon_1+\epsilon_2)(\epsilon_1+\epsilon_3)(\epsilon_2+\epsilon_3)}{\epsilon_1\epsilon_2\epsilon_3\epsilon_4}}\;,\\
    &\mathcal{F}(p,v_1-v_2,\{\epsilon_k\})=\frac{(v_1-v_2)(\epsilon_1+\epsilon_2)(\epsilon_1+\epsilon_3)(\epsilon_2+\epsilon_3)p}{\epsilon_1\epsilon_2\epsilon_3\epsilon_4(1-p)^2}\;,
\end{align}
where
\begin{equation}
    M(x,t):=\prod_{k=1}^{\infty}\frac{1}{(1-xt^k)^k}
\end{equation}
is the MacMahon function, and $M(t)=M(1,t)$.

\paragraph{Elliptic invariants} For the elliptic genus, we have
\begin{align}
    Z_\text{1-loop}=&\left(\frac{-2\pi\eta(\tau)^3\theta_1(\tau,\epsilon_1+\epsilon_2)\theta_1(\tau,\epsilon_1+\epsilon_3)\theta_1(\tau,\epsilon_2+\epsilon_3)}{\theta_1(\tau,-\epsilon_1)\theta_1(\tau,-\epsilon_2)\theta_1(\tau,-\epsilon_3)\theta_1(\tau,-\epsilon_4)}\right)^N\left(\prod_{i=1}^N\frac{-\theta_1(\tau,v_2-u_i)}{\theta_1(\tau,u_i-v_1)}\text{d}u_i\right)\nonumber\\
    &\left(\prod_{i\neq j}^N\frac{\theta_1(\tau,u_j-u_i)\theta_1(\tau,u_j-u_i+\epsilon_1+\epsilon_2)\theta_1(\tau,u_j-u_i+\epsilon_1+\epsilon_3)\theta_1(\tau,u_j-u_i+\epsilon_2+\epsilon_3)}{\theta_1(\tau,u_j-u_i-\epsilon_1)\theta_1(\tau,u_j-u_i-\epsilon_2)\theta_1(\tau,u_j-u_i-\epsilon_3)\theta_1(\tau,u_j-u_i-\epsilon_4)}\right)\;.
\end{align}
Notice that $-\theta_1(\tau,v_2-u_i)=\theta_1(\tau,u_i-v_2)$. Indeed, under the transformation of $u_i\rightarrow u_i+a+b\tau$ with $a,b\in\mathbb{Z}$ for any $u_i$, the integrand would get an extra phase $\text{e}^{-2\pi\text{i}b(v_1-v_2)}$. Therefore, we need to have $v_1-v_2\in\mathbb{Z}$.

\subsection{Orbifolds: \texorpdfstring{$\mathbb{C}^2\times\mathbb{C}^2/\mathbb{Z}_n$}{C2 x C2/Zn}}\label{C2C2Zn}
Let us now consider a special family of the $\mathbb{C}^4$ orbifolds, namely $\mathbb{C}^2\times\mathbb{C}^2/\mathbb{Z}_n$. The toric diagram and the quiver are given as
\begin{equation}
\begin{matrix}
    \includegraphics[trim=0 10 0 0,width=16cm]{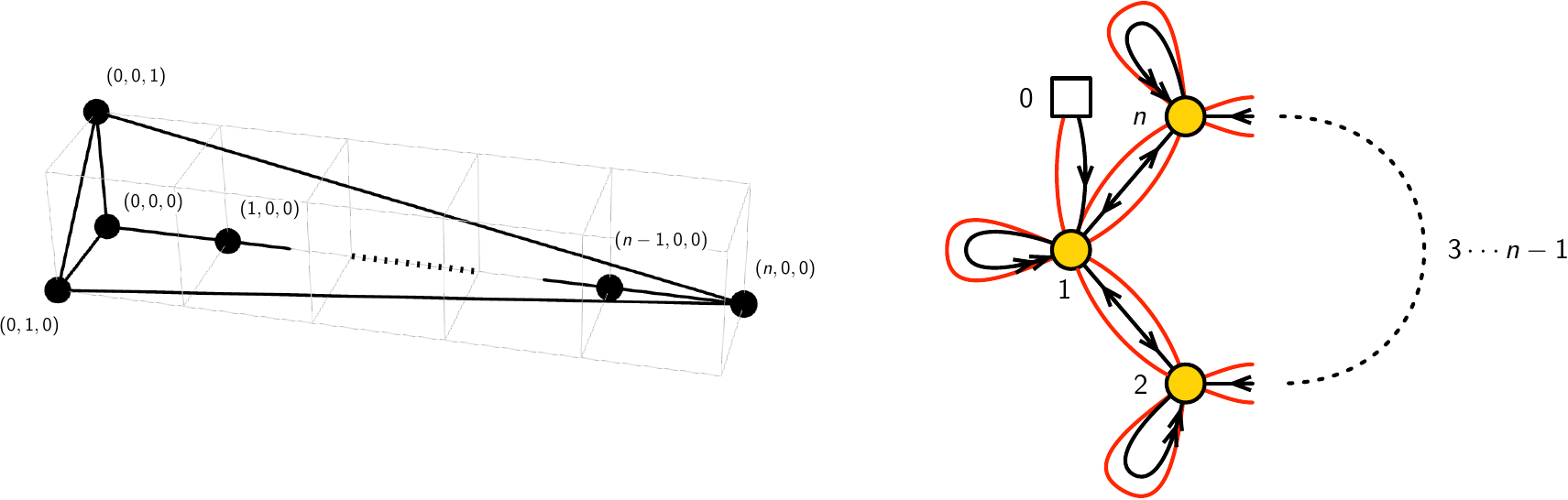}.
\end{matrix}
\end{equation}
\\
When $n=2$, there are only two gauge nodes, and there are two pairs of opposite chirals and four Fermis connecting the two nodes. The weights of the adjoint chirals (resp.~Fermi) for each node are taken to be $\epsilon_{3,4}$ (resp.~$-\epsilon_1-\epsilon_2$). For nodes $a$ and $a+1$, the chiral $a\rightarrow a+1$ (resp.~$a+1\rightarrow a$) connecting them has weight $\epsilon_1$ (resp.~$\epsilon_2$) while the Fermis have weights $-\epsilon_1-\epsilon_3$ and $-\epsilon_2-\epsilon_3$. Here, $a=n+1$ is understood as $a\equiv1~(\text{mod }n)$. The edges connected to the framing node are assigned the same weights as in the $\mathbb{C}^4$ case.

For the dimension vector $\bm{N}=(N_a)$, the integrand is then
\begin{align}
    Z_\text{1-loop}
    &= q_3^{\sum\limits_{a\in Q_0}\frac{(N_a-N_{a+1})^2}{2}}q_4^{\sum\limits_{a\in Q_0}N_a^2}\left(\frac{1-q_1q_2}{(1-q_3)(1-q_4)}\right)^{\sum\limits_{a\in Q_0}N_a}\left(\sqrt{\frac{w_1}{w_2}}\right)^{N_1}\left(\prod_{a\in Q_0}\prod_{i=1}^{N_a}\frac{\text{d}x^{(a)}_i}{x^{(a)}_i}\right)\nonumber\\
    &\left(\prod_{i=1}^{N_1}\frac{\left(x^{(1)}_i-w_2\right)}{\left(x^{(1)}_i-w_1\right)}\right)\left(\prod_{a\in Q_0}\prod_{i\neq j}^{N_a}\frac{\left(x^{(a)}_i-x^{(a)}_j\right)\left(x^{(a)}_j-x^{(a)}_iq_1q_2\right)}{\left(x^{(a)}_i-x^{(a)}_jq_3\right)\left(x^{(a)}_i-x^{(a)}_jq_4\right)}\right)\nonumber\\
    &\left(\prod_{a\in Q_0}\prod_{i=1}^{N_a}\prod_{j=1}^{N_{a+1}}\frac{\left(x^{(a)}_i-x^{(a+1)}_jq_2q_3\right)\left(x^{(a+1)}_j-x^{(a)}_iq_1q_3\right)}{\left(x^{(a+1)}_j-x^{(a)}_iq_1\right)\left(x^{(a)}_i-x^{(a+1)}q_2\right)}\right)\;.
\end{align}
Again, we have $n+1\equiv1$. As an illustration, let us list some indices at low ranks for $n=2$:
\begin{itemize}
    \item Level 1:
    \begin{itemize}
        \item crystal with a single atom of colour $p_1$:
        \begin{equation}
        \mathcal{I}_1=\frac{(1-q_1q_2)q_3q_4\left(\sqrt{\mu}-1/\sqrt{\mu}\right)}{(1-q_3)(1-q_4)}\;.
    \end{equation}
    \end{itemize}
    \item Level 2:
    \item crystal with an atom of colour $p_1$ and an atom of colour $p_2$:
    \begin{equation}
        \mathcal{I}_{2,(11)}=\sum_{i=1}^2\frac{(-1)^{i+1}q_4(1-q_1q_2)(1-q_iq_1q_3)(1-q_iq_2q_3)(q_i-q_1q_3)(q_i-q_2q_3)}{q_iq_3(1-q_3)(1-q_4)(1-q_iq_1)(1-q_iq_2)(q_1-q_2)}\mathcal{I}_1 \;.
    \end{equation}
    \item crystal with two atoms of colour $p_1$:
    \begin{equation}
        \mathcal{I}_{2,(2,0)}=\sum_{i=3}^4\frac{(-1)^{i+1}q_3^3q_4^3(1-q_1q_2)(1-q_i)(1-q_iq_1q_2)(q_i-q_1q_2)\left(q_i\sqrt{\mu}-1/\sqrt{\mu}\right)}{q_i(1-q_3)(1-q_4)(1-q_iq_3)(1-q_iq_4)(q_3-q_4)}\mathcal{I}_1 \;.
    \end{equation}
\end{itemize}

\paragraph{Rational limit} In the rational limit, the integrand is
\begin{align}
    Z_\text{1-loop}=&\left(-\frac{\epsilon_1+\epsilon_2}{\epsilon_3\epsilon_4}\right)^{\sum\limits_{a\in Q_0}N_a}\left(\prod_{a\in Q_0}\prod_{i=1}^{N_a}\text{d}u^{(a)}_i\right)\left(\prod_{i=1}^{N_1}\frac{u^{(1)}_i-v_2}{u^{(1)}_i-v_1}\right)\nonumber\\
    &\left(\prod_{a\in Q_0}\prod_{i\neq j}^{N_a}\frac{\left(u^{(a)}_i-u^{(a)}_j\right)\left(u^{(a)}_j-u^{(a)}_i-\epsilon_1-\epsilon_2\right)}{\left(u^{(a)}_i-u^{(a)}_j-\epsilon_3\right)\left(u^{(a)}_i-u^{(a)}_j-\epsilon_4\right)}\right)\nonumber\\
    &\left(\prod_{a\in Q_0}\prod_{i=1}^{N_a}\prod_{j=1}^{N_{a+1}}\frac{\left(u^{(a)}_i-u^{(a+1)}_j-\epsilon_2-\epsilon_3\right)\left(u^{(a+1)}_j-u^{(a)}_i-\epsilon_1-\epsilon_3\right)}{\left(u^{(a+1)}_j-u^{(a)}_i-\epsilon_1\right)\left(u^{(a)}_i-u^{(a+1)}_j-\epsilon_2\right)}\right) \;.
\end{align}
In fact, it was conjectured in \cite{Szabo:2023ixw} that
\begin{align}
    &Z_\text{mat}=\text{PE}[\mathcal{F}]\left(p,\{p_{[a,b]}\},v_1-v_2,\{\epsilon_k\}\right)\nonumber\\
    &\qquad=M(p)^{\frac{n(v_1-v_2)(\epsilon_1+\epsilon_2)(\epsilon_1+\epsilon_3)(\epsilon_2+\epsilon_3)}{\epsilon_1\epsilon_2\epsilon_3\epsilon_4}+\left(n-\frac{1}{n}\right)\frac{(v_1-v_2)(\epsilon_1+\epsilon_2)}{\epsilon_1\epsilon_2}}\prod_{1<a\leq b\leq n}\widetilde{M}\left(p_{[a,b]},p\right)^{\frac{(v_1-v_2)(\epsilon_1+\epsilon_2)}{\epsilon_3\epsilon_4}},\\
    &\mathcal{F}\left(p,\{p_{[a,b]}\},v_1-v_2,\{\epsilon_k\}\right)=\left(\frac{n(\epsilon_1+\epsilon_3)(\epsilon_2+\epsilon_3)}{\epsilon_1\epsilon_2\epsilon_3\epsilon_4}+\frac{n-1/n}{\epsilon_1\epsilon_2}+\sum_{1<a\leq b\leq n}\frac{p_{[a,b]}+1/p_{[a,b]}}{\epsilon_3\epsilon_4}\right)\nonumber\\
    &\qquad\qquad\qquad\qquad\qquad\qquad\quad\frac{(v_1-v_2)(\epsilon_1+\epsilon_2)p}{(1-p)^2} \;,
\end{align}
where
\begin{equation}
    \widetilde{M}(x,t)=M(x,t)M\left(x^{-1},t\right) \;.
\end{equation}
Here, we have introduced the notations $p_{[a,b]}:=(-1)^{a-b+1}p_ap_{a+1}\dots p_b$ (notice the sign). In the limit $\epsilon_4/(v_1-v_2)\rightarrow0$, this recovers the partition function for the threefold $\mathbb{C}\times\mathbb{C}^2/\mathbb{Z}_n$.

\paragraph{Elliptic invariants} For the elliptic genus, we have
\begin{align}
    Z_\text{1-loop}=&\left(\frac{-2\pi\eta(\tau)^3\theta_1(\tau,\epsilon_1+\epsilon_2)}{\theta_1(\tau,-\epsilon_3)\theta_1(\tau,-\epsilon_4)}\right)^{\sum\limits_{a\in Q_0}N_a}\left(\prod_{a\in Q_0}\prod_{i=1}^{N_a}\text{d}u_a\right)\left(\prod_{i=1}^{N_1}\frac{-\theta_1\left(\tau,v_2-u^{(1)}_i\right)}{\theta_1\left(\tau,u^{(1)}_i-v_1\right)}\right)\nonumber\\
    &\left(\prod_{a\in Q_0}\prod_{i\neq j}^{N_a}\frac{\theta_1\left(\tau,u^{(a)}_i-u^{(a)}_j\right)\theta_1\left(\tau,u^{(a)}_i-u^{(a)}_j+\epsilon_1+\epsilon_2\right)}{\theta_1\left(\tau,u^{(a)}_i-u^{(a)}_j-\epsilon_3\right)\theta_1\left(u^{(a)}_i-u^{(a)}_j-\epsilon_4\right)}\right)\nonumber\\
    &\left(\prod_{a\in Q_0}\prod_{i=1}^{N_a}\prod_{j=1}^{N_a+1}\frac{\theta_1\left(u^{(a+1)}_j-u^{(a)}_i+\epsilon_2+\epsilon_3\right)\theta_1\left(u^{(a)}_i-u^{(a+1)}_j+\epsilon_1+\epsilon_3\right)}{\theta_1\left(u^{(a+1)}_j-u^{(a)}_i-\epsilon_1\right)\theta_1\left(u^{(a)}_i-u^{(a+1)}_j-\epsilon_2\right)}\right)\;.
\end{align}
Again, the gauge anomaly would require $v_1-v_2\in\mathbb{Z}$.

\subsubsection{Wall Crossing of \texorpdfstring{$\mathbb{C}^2\times\mathbb{C}^2/\mathbb{Z}_2$}{C2xC2/Z2}}\label{wallcrossingC2C2Z2}
Let us consider the wall crossing of the specific example with $n=2$. We shall focus on the chambers that are induced by the cyclic chambers of the corresponding 4d quivers. For $\mathbb{C}\times\mathbb{C}^2/\mathbb{Z}_2$, the cyclic chambers have the structure
\begin{equation}
\begin{matrix}
    \includegraphics[width=14cm]{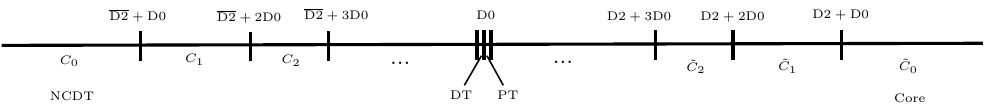}.\label{cyclicchambers}
\end{matrix}
\end{equation}
For the chambers $C_K$, the crystals are coloured plane partitions with $K$ semi-infinite faces ``peeled off'':
\begin{equation}
\begin{matrix}
    \includegraphics[width=14cm]{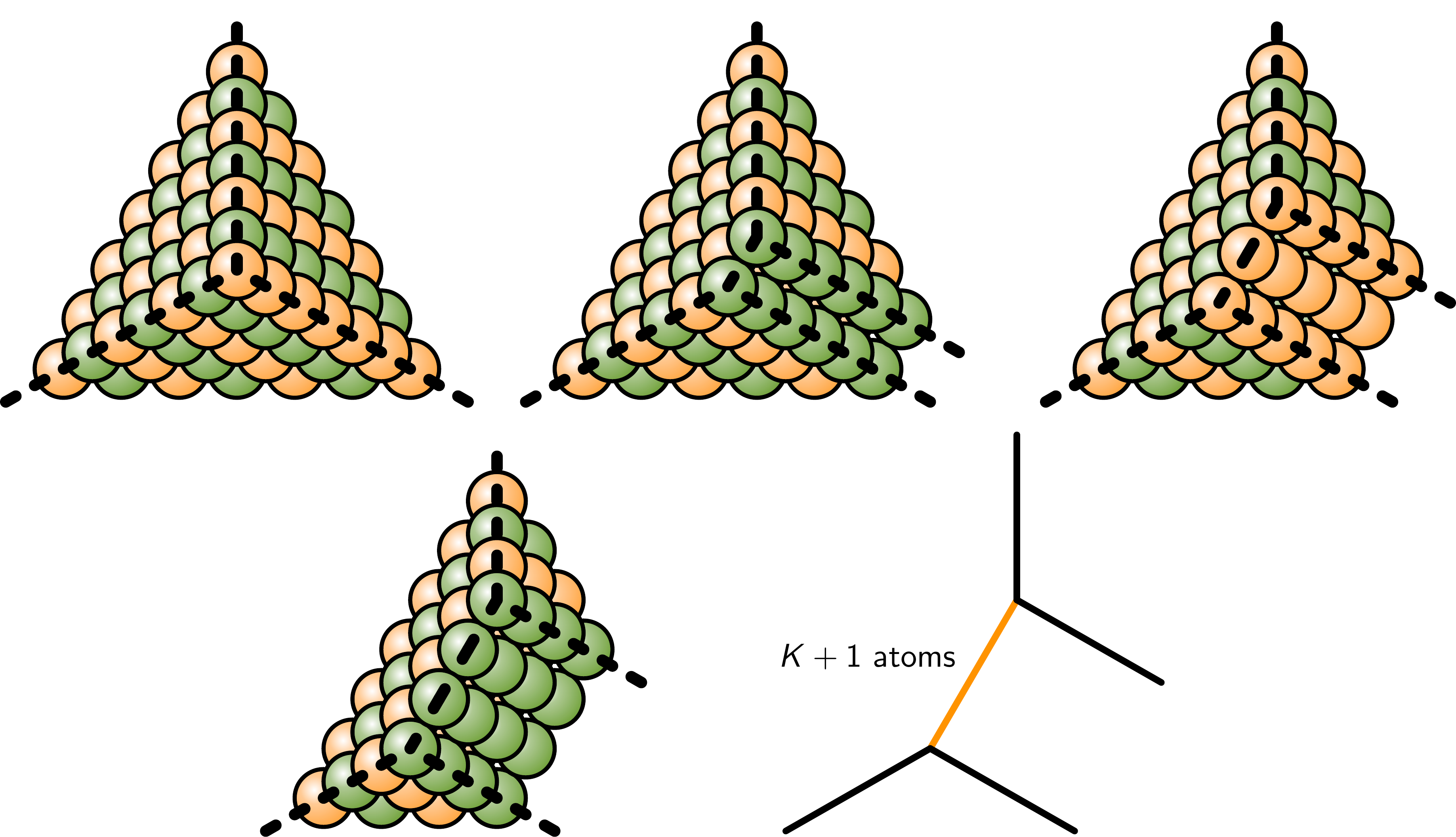}.\label{cpp}
\end{matrix}
\end{equation}
For the chambers $\widetilde{C}_K$, the crystals are of the Toblerone shape \cite{Bao:2022oyn}:
\begin{equation}
\begin{matrix}
    \includegraphics[width=10cm]{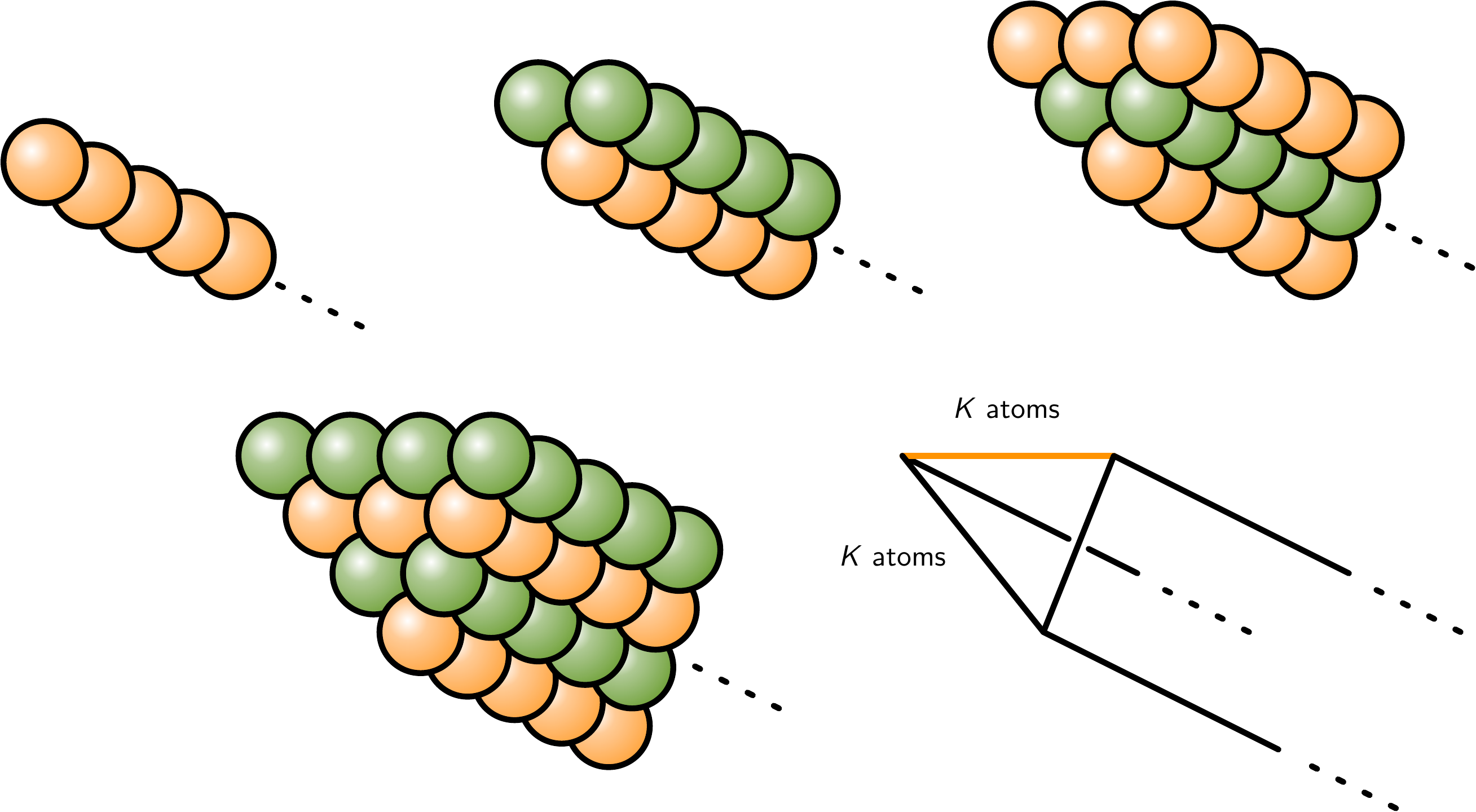}.\label{toblerone}
\end{matrix}
\end{equation}
Here, the orange lines indicate the top rows in the crystals. Notice that the first one is for $K=1$ in this figure as $\widetilde{C}_{0}$ trivially has $Z_\text{BPS}=1$.

For the chambers of $\mathbb{C}^2\times\mathbb{C}^2/\mathbb{Z}_2$ that are induced by the above chambers of $\mathbb{C}\times\mathbb{C}^2/\mathbb{Z}_2$, we shall use the same notations $C_K$ and $\widetilde{C}_K$. Moreover, it is also clear how the framing of the quiver would change:
\begin{equation}
\begin{matrix}
    \includegraphics[width=16cm]{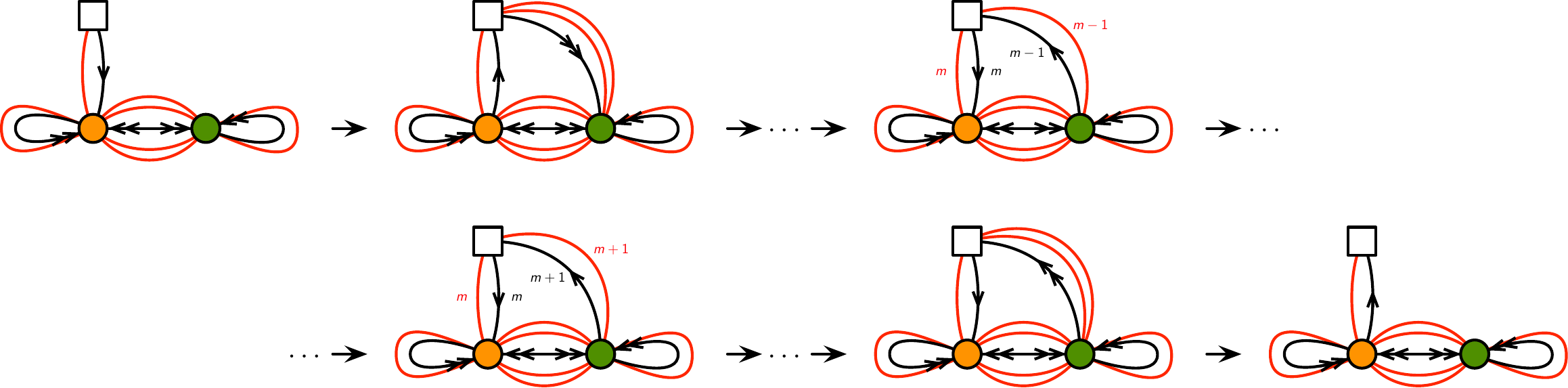}.
\end{matrix}
\end{equation}
For convenience, let us swap the labels of the two gauge nodes every time we cross a wall of marginal stability. In other words, $a=1$ always refers to the gauge node with incoming chirals from the framing node.

For the $C_K$ chamber, the chirals connecting the framing node and the gauge node 1 (resp.~2) have weights $v_1+K\epsilon_1$, $v_1+(K-1)\epsilon_1+\epsilon_2$, $\dots$, $v_1+K\epsilon_2$ (resp.~$v_2+K\epsilon_1+\epsilon_2$, $v_2+(K-1)\epsilon_1+2\epsilon_2$, $\dots$, $v_2+\epsilon_1+K\epsilon_2$ when existing) while the accompanied Fermis have weights $-v_2-K\epsilon_1$, $-v_2-(K-1)\epsilon_1-\epsilon_2$, $\dots$, $-v_2-K\epsilon_2$ (resp.~$v_1+K\epsilon_1+\epsilon_2$, $v_1+(K-1)\epsilon_1+2\epsilon_2$, $\dots$, $v_1+\epsilon_1+K\epsilon_2$ when existing). Therefore, in the integrand, we only need to change the factors coming from the framing to
\begin{align}
    &\left(\prod_{i=1}^{N_1}\frac{\left(u^{(1)}_i-v_2-K\epsilon_1\right)\dots\left(u^{(1)}_i-v_2-K\epsilon_2\right)}{\left(u^{(1)}_i-v_1-K\epsilon_1\right)\dots\left(u^{(1)}_i-v_1-K\epsilon_2\right)}\right)\nonumber\\
    &\left(\prod_{i=1}^{N_2}\frac{\left(v_1+K\epsilon_1+\epsilon_2-u^{(2)}_i\right)\dots\left(v_1+\epsilon_1+K\epsilon_2-u^{(2)}_i\right)}{\left(v_2+K\epsilon_1+\epsilon_2-u^{(2)}_i\right)\dots\left(v_2+\epsilon_1+K\epsilon_2-u^{(2)}_i\right)}\right)\;,
\end{align}
where we have used the equivariant version for simplicity. We conjecture that the equivariant partition function is
\begin{align}
    &Z_\text{mat}=\text{PE}[\mathcal{F}]\left(p_{(K)},p_{(K,[2,2])},v_1-v_2,\{\epsilon_k\}\right)\nonumber\\
    &\qquad=M\left(p_{(K)}\right)^{\frac{2(v_1-v_2)(\epsilon_1+\epsilon_2)(\epsilon_1+\epsilon_3)(\epsilon_2+\epsilon_3)}{\epsilon_1\epsilon_2\epsilon_3\epsilon_4}+\frac{3(v_1-v_2)(\epsilon_1+\epsilon_2)}{2\epsilon_1\epsilon_2}}\nonumber\\
    &\qquad\quad~\left(M\left(p_{(K,[2,2])},p_{(K)}\right)\prod_{k=K+1}^{\infty}\frac{1}{\left(1-p_{(K,[2,2])}^{-1}p_{(K)}^k\right)^k}\right)^{\frac{(v_1-v_2)(\epsilon_1+\epsilon_2)}{\epsilon_3\epsilon_4}}\;,\\
    &\mathcal{F}\left(p_{(K)},p_{(K,[2,2])},v_1-v_2,\{\epsilon_k\}\right)\nonumber\\
    &=\frac{(v_1-v_2)(\epsilon_1+\epsilon_2)p_{(K)}}{(1-p_{(K)})^2}\left(\frac{2(\epsilon_1+\epsilon_3)(\epsilon_2+\epsilon_3)}{\epsilon_1\epsilon_2\epsilon_3\epsilon_4}+\frac{3}{2\epsilon_1\epsilon_2}\right)\nonumber\\
    &+\frac{(v_1-v_2)(\epsilon_1+\epsilon_2)}{\epsilon_3\epsilon_4}\left(\frac{\left(p_{(K,[2,2])}+1/p_{(K,[2,2])}\right)p_{(K)}}{\left(1-p_{(K)}\right)^2}-\frac{1}{p_{(K,[2,2])}}\sum_{k=1}^Kkp_{(K)}^K\right)\;,
\end{align}
where $p_{(K)}=p_1p_2$ and $p_{(K,[2,2])}=-p_1^Kp_2^{K+1}$. The crystals are the 4d versions of \eqref{cpp}, where one ``peels off'' a 3d subcrystal every time one crosses a wall from $C_K$ to $C_{K+1}$.

For the $\widetilde{C}_K$ chamber, the chirals connecting the framing node and the gauge node 2 (resp.~1) have weights $v_2+K\epsilon_1$, $v_2+(K-1)\epsilon_1+\epsilon_2$, $\dots$, $v_2+K\epsilon_2$ (resp.~$v_1+(K-1)\epsilon_1$, $v_1+(K-2)\epsilon_1+\epsilon_2$, $\dots$, $v_1+(K-1)\epsilon_2$ when existing) while the accompanied Fermis have weights $v_1+K\epsilon_1$, $v_1+(K-1)\epsilon_1+\epsilon_2$, $\dots$, $v_1+K\epsilon_2$ (resp.~$-v_2-(K-1)\epsilon_1$, $-v_2-(K-2)\epsilon_1-\epsilon_2$, $\dots$, $-v_2-(K-1)\epsilon_2$ when existing). Therefore, in the integrand, we only need to change the factors coming from the framing to
\begin{align}
    &\left(\prod_{i=1}^{N_1}\frac{\left(u^{(1)}_i-v_2-(K-1)\epsilon_1\right)\dots\left(u^{(1)}_i-v_2-(K-1)\epsilon_2\right)}{\left(u^{(1)}_i-v_1-(K-1)\epsilon_1\right)\dots\left(u^{(1)}_i-v_1-(K-1)\epsilon_2\right)}\right)\nonumber\\
    &\left(\prod_{i=1}^{N_2}\frac{\left(v_1+K\epsilon_1-u^{(2)}_i\right)\dots\left(v_1+K\epsilon_2-u^{(2)}_i\right)}{\left(v_2+K\epsilon_1-u^{(2)}_i\right)\dots\left(v_2+K\epsilon_2-u^{(2)}_i\right)}\right)\;,
\end{align}
where we have used the equivariant version for simplicity. We conjecture that the equivariant partition function is
\begin{align}
    &Z_\text{mat}=\text{PE}[\mathcal{F}]\left(\widetilde{p}_{(K)},\widetilde{p}_{(K,[2,2])},v_1-v_2,\{\epsilon_k\}\right)=\left(\prod_{k=1}^K\frac{1}{\left(1-\widetilde{p}_{(K,[2,2])}\widetilde{p}_{(K)}^k\right)^k}\right)^{\frac{(v_1-v_2)(\epsilon_1+\epsilon_2)}{\epsilon_3\epsilon_4}}\;,\\
    &\mathcal{F}\left(\widetilde{p}_{(K)},\widetilde{p}_{(K,[2,2])},v_1-v_2,\{\epsilon_k\}\right)=\frac{\widetilde{p}_{(K,[2,2])}(v_1-v_2)(\epsilon_1+\epsilon_2)}{\epsilon_3\epsilon_4}\left(\sum_{k=1}^K\sum_{l=1}^k(\sqrt{\mu})^{2l-k-1}\widetilde{p}_{(K)}^K\right)\;,
\end{align}
where $\widetilde{p}_{(K)}=p_1^{-1}p_2^{-1}$ and $\widetilde{p}_{(K,[2,2])}=-p_1^{K+1}p_2^K$. The crystals are the 4d versions of \eqref{toblerone}, where one adds a ``layer'' of a 3d subcrystal every time one crosses a wall from $\widetilde{C}_K$ to $\widetilde{C}_{K+1}$.

Let us also make a comment on the elliptic invariants. For the chamber $C_K$, the shift of $u^{(1)}_i\rightarrow u^{(1)}_i+a+b\tau$ with $a,b\in\mathbb{Z}$ would yield an extra phase $\text{e}^{-2\pi\text{i}(K+1)(v_1-v_2)}$, and hence $(K+1)(v_1-v_2)\in\mathbb{Z}$. On the other hand, the transformation of $u^{(2)}_i$ would lead to $K(v_1-v_2)\in\mathbb{Z}$. As $K+1$ and $K$ are coprime, we should again have $v_1-v_2\in\mathbb{Z}$. Likewise, for the chamber $\widetilde{C}_K$, the transformations of $u^{(1)}_i$ and $u^{(2)}_i$ would give rise to $K(v_1-v_2)\in\mathbb{Z}$ and $(K+1)(v_1-v_2)\in\mathbb{Z}$. Therefore, we should have $v_1-v_2\in\mathbb{Z}$.

\subsection{\texorpdfstring{Conifold$\times\mathbb{C}$}{Conifold x C}}\label{conc}
Another typical example would be the conifold$\times\mathbb{C}$. The toric diagram and the quiver are
\begin{equation}
\begin{matrix}
    \includegraphics[trim=0 10 0 -10,width=11cm]{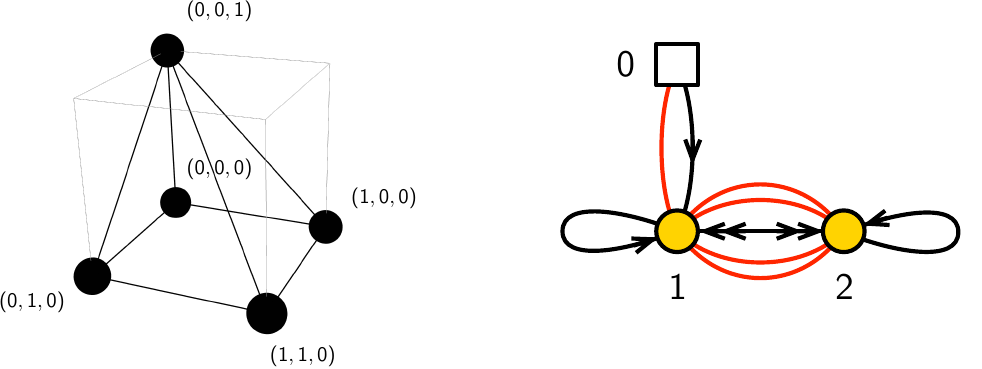}.
\end{matrix}
\end{equation}
The weights of the edges are
\begin{equation}
    \begin{tabular}{c|c|c|c|c|c|c|c|c|c}
$\chi_{11}$ & $\chi_{12,1}$ & $\chi_{12,2}$ & $\chi_{21,1}$ & $\chi_{21,2}$ & $\chi_{22}$ & $\Lambda_{12,1}$ & $\Lambda_{12,2}$ & $\Lambda_{21,1}$ & $\Lambda_{21,2}$ \\ \hline\hline
$\epsilon_4$ & $\epsilon_1$ & $-\epsilon_1$ & $\epsilon_2$ & $-\epsilon_2-\epsilon_4$ & $\epsilon_4$ & $-\epsilon_1+\epsilon_4$ & $\epsilon_1+\epsilon_4$ & $\epsilon_2+\epsilon_4$ & $-\epsilon_2$ 
\end{tabular},\label{conifoldCweights}
\end{equation}
which can be directly obtained from the dimensional reduction of the conifold case. The weights of the chiral and the Fermi connected to the framing node are still $v_1$ and $-v_2$ respectively.

For the dimension vector $\bm{N}=(N_1,N_2)$, the integrand is then
\begin{align}
    Z_\text{1-loop}=&q_4^{\frac{(N_1-N_2)^2}{2}}\left(\frac{1}{1-q_4}\right)^{N_1+N_2}\left(\sqrt{\frac{w_1}{w_2}}\right)^{N_1}\left(\prod_{i=1}^{N_1}\frac{\text{d}x^{(1)}_i}{x^{(1)}_i}\right)\left(\prod_{i=1}^{N_2}\frac{\text{d}x^{(2)}_i}{x^{(2)}_i}\right)\nonumber\\
    &\left(\prod_{i=1}^{N_1}\frac{w_2-x^{(1)}_i}{x^{(1)}_i-w_1}\right)\left(\prod_{i\neq j}^{N_1}\frac{x^{(1)}_i-x^{(1)}_j}{x^{(1)}_i-x^{(1)}_jq_4}\right)\left(\prod_{i\neq j}^{N_2}\frac{x^{(2)}_i-x^{(2)}_j}{x^{(2)}_i-x^{(2)}_jq_4}\right)\nonumber\\
    &\left(\prod_{i=1}^{N_1}\prod_{j=1}^{N_2}\frac{\left(x^{(1)}_i-x^{(2)}_jq_1^{-1}q_3^{-1}\right)\left(x^{(1)}_i-x^{(2)}_jq_2^{-1}\right)\left(x^{(2)}_j-x^{(1)}_iq_1q_4^{-1}\right)\left(x^{(2)}_j-x^{(1)}_iq_1^{-1}q_4^{-1}\right)}{\left(x^{(1)}_i-x^{(2)}_jq_2\right)\left(x^{(1)}_i-x^{(2)}_jq_1q_3\right)\left(x^{(2)}_j-x^{(1)}_iq_1\right)\left(x^{(2)}_j-x^{(1)}_iq_1^{-1}\right)}\right)\;.
\end{align}
As an illustration, let us list some indices at low ranks:
\begin{itemize}
    \item Level 1:
    \begin{itemize}
        \item crystal with a single atom of colour $p_1$:
        \begin{equation}
        \mathcal{I}_1=\frac{\sqrt{q_4}\left(\sqrt{\mu}-1/\sqrt{\mu}\right)}{(q_4-1)}\;.
    \end{equation}
    \end{itemize}
    \item Level 2:
    \item crystal with an atom of colour $p_1$ and an atom of colour $p_2$:
    \begin{equation}
        \mathcal{I}_{2,(11)}=\sum_{\varsigma=\pm1}\frac{\varsigma\left(1-q_1^{\varsigma}q_1^{-1}q_3^{-1}\right)\left(1-q_1^{\varsigma}q_2^{-1}\right)\left(q_1^{\varsigma}-q_1q_4^{-1}\right)\left(q_1^{\varsigma}-q_1^{-1}q_4^{-1}\right)}{\sqrt{q_4}q_1^{\varsigma}(1-q_4)\left(q_1-q_1^{-1}\right)\left(1-q_1^{\varsigma}q_2\right)\left(1-q_1^{\varsigma}q_1q_3\right)}\mathcal{I}_1\;,
    \end{equation}
    \item crystal with two atoms of colour $p_1$:
    \begin{equation}
        \mathcal{I}_{2,(2,0)}=-\frac{\sqrt{q_4}\left(q_4\sqrt{\mu}-1/\sqrt{\mu}\right)}{1-q_4}\mathcal{I}_1\;.
    \end{equation}
\end{itemize}

\paragraph{Rational limit} In the rational limit, the integrand is
\begin{align}
    &Z_\text{1-loop}\nonumber\\
    =&\left(-\frac{1}{\epsilon_4}\right)^{N_1+N_2}\left(\prod_{a=1}^2\prod_{i=1}^{N_a}\text{d}u^{(a)}_i\right)\left(\prod_{i=1}^{N_1}\frac{u^{(1)}_i-v_2}{u^{(1)}_i-v_1}\right)\left(\prod_{a=1}^2\prod_{i\neq j}^{N_a}\frac{\left(u^{(a)}_i-u^{(a)}_j\right)}{\left(u^{(a)}_i-u^{(a)}_j-\epsilon_4\right)}\right)\nonumber\\
    &\left(\prod_{i=1}^{N_1}\prod_{j=1}^{N_2}\frac{\left(u^{(1)}_i-u^{(2)}_j+\epsilon_1+\epsilon_3\right)\left(u^{(1)}_i-u^{(2)}_j+\epsilon_2\right)\left(u^{(2)}_j-u^{(1)}_i-\epsilon_1+\epsilon_4\right)\left(u^{(2)}_j-u^{(1)}_i+\epsilon_1+\epsilon_4\right)}{\left(u^{(1)}_i-u^{(2)}_j-\epsilon_2\right)\left(u^{(1)}_i-u^{(2)}_j-\epsilon_1-\epsilon_3\right)\left(u^{(2)}_j-u^{(1)}_i-\epsilon_1\right)\left(u^{(2)}_j-u^{(1)}_i+\epsilon_1\right)}\right)\;.
\end{align}
We conjecture that the equivariant partition function is
\begin{align}
    &Z_\text{mat}=\text{PE}[\mathcal{F}]\left(p',p_{[2,2]},v_1-v_2,\{\epsilon_k\}\right)\nonumber\\
    &\qquad=M(p')^{\frac{(v_1-v_2)(\epsilon_1+\epsilon_2)(2\epsilon_1+\epsilon_3)(2\epsilon_1+\epsilon_4)}{2\epsilon_1\epsilon_3\epsilon_4(\epsilon_1-\epsilon_2)}+\frac{(v_1-v_2)\epsilon_3(\epsilon_1-\epsilon_2)(2\epsilon_1-\epsilon_4)}{2\epsilon_1\epsilon_4(\epsilon_1+\epsilon_2)(2\epsilon_1+\epsilon_3)}}\widetilde{M}\left(p_{[2,2]},p'\right)^{\frac{(v_1-v_2)}{\epsilon_4}},\\
    &\mathcal{F}\left(p',p_{[2,2]},v_1-v_2,\{\epsilon_k\}\right)=\left(\frac{(\epsilon_1+\epsilon_2)(2\epsilon_1+\epsilon_3)(2\epsilon_1+\epsilon_4)}{2\epsilon_1\epsilon_3\epsilon_4(\epsilon_1-\epsilon_2)}+\frac{\epsilon_3(\epsilon_1-\epsilon_2)(2\epsilon_1-\epsilon_4)}{2\epsilon_1\epsilon_4(\epsilon_1+\epsilon_2)(2\epsilon_1+\epsilon_3)}\right.\nonumber\\
    &\qquad\qquad\qquad\qquad\qquad\qquad\quad\left.+\frac{p_{[2,2]}+1/p_{[2,2]}}{\epsilon_4}\right)\frac{(v_1-v_2)p'}{(1-p')^2}\;,
\end{align}
where $p':=-p_0p_1$ and $p_{[2,2]}:=-p_1$ (notice the sign). In the limit $\epsilon_4/(v_1-v_2)\rightarrow0$, this recovers the partition function for the conifold.

\paragraph{Elliptic invariants} The integrand for the elliptic genus is straightforward to write down. Hence, we shall omit the full expression here. Again, due to the factor
\begin{equation}
    \left(\prod_{i=1}^{N_1}\frac{-\theta_1\left(\tau,v_2-u^{(1)}_i\right)}{\theta_1\left(\tau,u^{(1)}_i-v_1\right)}\right),
\end{equation}
the gauge anomaly would require $v_1-v_2\in\mathbb{Z}$.

\subsubsection{Wall Crossing}\label{wallcrossingconc}
Let us now consider the wall crossing, focusing on the chambers that are induced by the cyclic chambers of the corresponding 4d quivers \cite{young2009computing}. For the conifold, the cyclic chambers have the same structure as given in \eqref{cyclicchambers}. For the infinite chambers $C_K$, the crystals look like
\begin{equation}
\begin{matrix}
    \includegraphics[width=14cm]{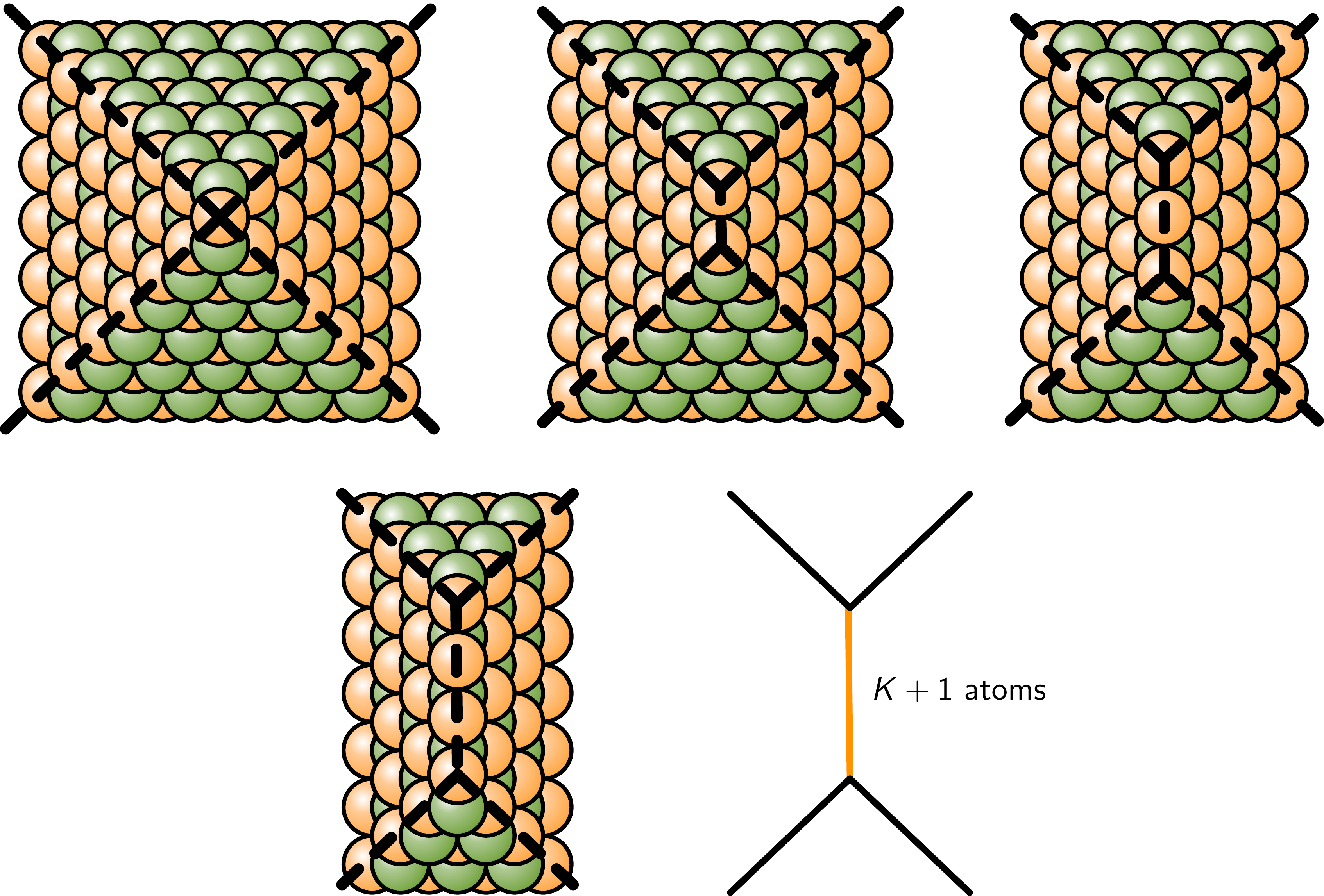}.\label{coninf}
\end{matrix}
\end{equation}
For the finite chambers $\widetilde{C}_K$, the crystals are
\begin{equation}
\begin{matrix}  
    \includegraphics[width=12cm]{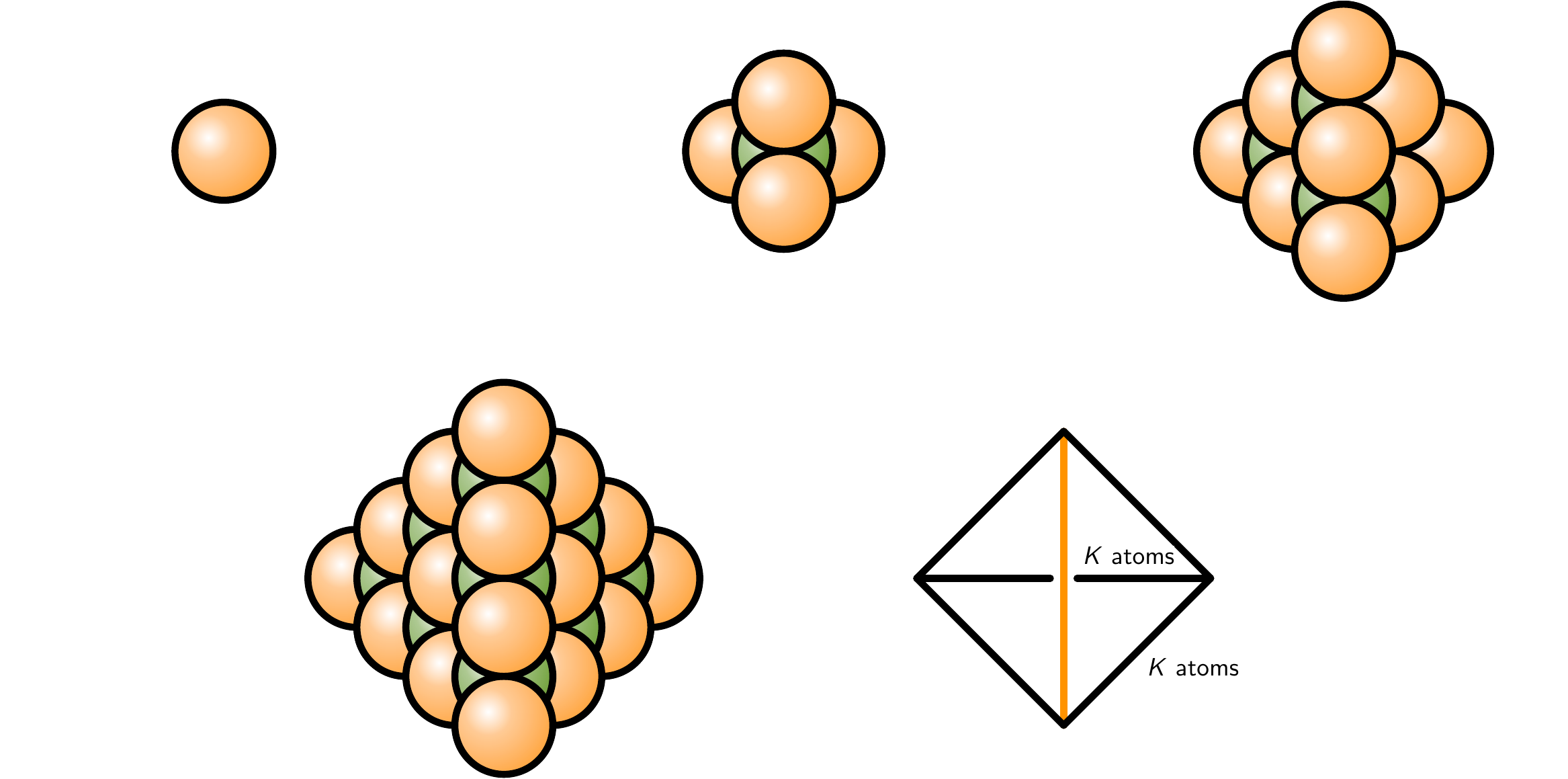}.\label{confin}
\end{matrix}    
\end{equation}
Here, the orange lines indicate the top rows in the crystals. Notice that the first one is for $K=1$ in this figure as $\widetilde{C}_0$ trivially has $Z_\text{BPS}=1$.

For the chambers of the conifold$\times\mathbb{C}$ that are induced by the above chamber of the conifold, we shall use the same notations $C_K$ and $\widetilde{C}_K$. Moreover, the framing of the quiver would change as in Fig.~\ref{wallcrossingex}, which is reproduced here:
\begin{equation}
\begin{matrix}
    \includegraphics[width=16cm]{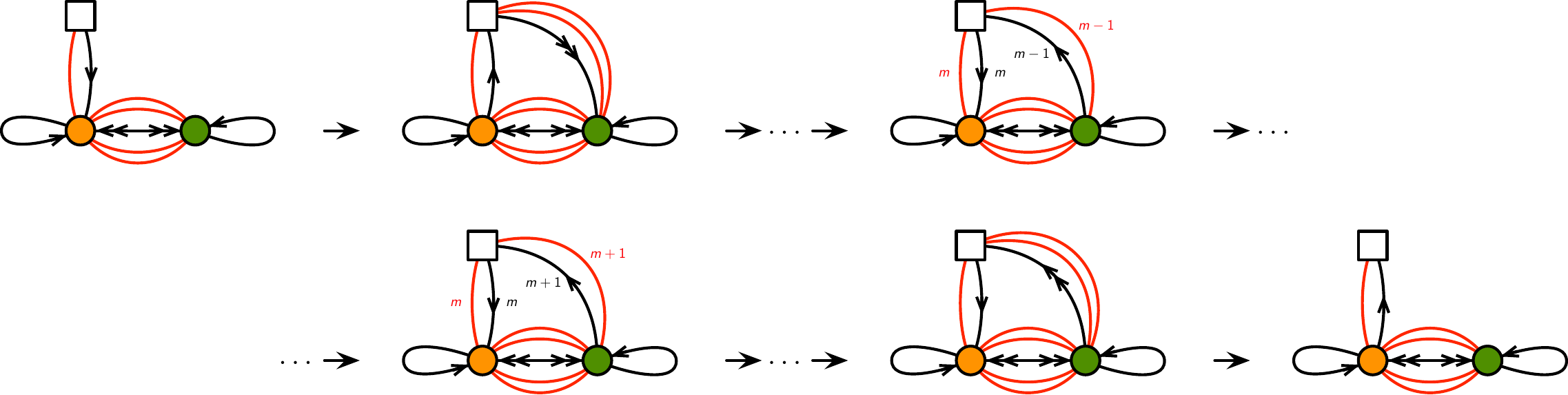}.
\end{matrix}
\end{equation}
\\
For convenience, we swap the colours and the labels of the two gauge nodes every time we cross a wall of marginal stability. In other words, $a=1$ always refers to the gauge node in red with incoming chirals from the framing node, and the weights of the edges remain the same as in \eqref{conifoldCweights}.

For the $C_K$ chamber, the chirals connecting the framing node and the gauge node 1 (resp.~2) have weights $v_1+K\epsilon_1+K\epsilon_2$, $v_1+(K-2)\epsilon_1+K\epsilon_2$, $\dots$, $v_1-K\epsilon_1+K\epsilon_2$ (resp.~$v_2+(K-1)\epsilon_1+K\epsilon_2$, $v_2+(K-3)\epsilon_1+K\epsilon_2$, $\dots$, $v_2-(K-1)\epsilon_1+K\epsilon_2$ when existing) while the accompanied Fermis have weights $-v_2-K\epsilon_1-K\epsilon_2$, $-v_2-(K-2)\epsilon_1-K\epsilon_2$, $\dots$, $-v_2+K\epsilon_1-K\epsilon_2$ (resp.~$v_1+(K-1)\epsilon_1+K\epsilon_2$, $v_1+(K-3)\epsilon_1+K\epsilon_2$, $\dots$, $v_1-(K-1)\epsilon_1+K\epsilon_2$ when existing). Therefore, in the integrand, we only need to change the factors coming from the framing to
\begin{align}
    &\left(\prod_{i=1}^{N_1}\frac{\left(u^{(1)}_i-v_2-K\epsilon_1-K\epsilon_2\right)\dots\left(u^{(1)}_i-v_2+K\epsilon_1-K\epsilon_2\right)}{\left(u^{(1)}_i-v_1-K\epsilon_1-K\epsilon_2\right)\dots\left(u^{(1)}_i-v_1+K\epsilon_1-K\epsilon_2\right)}\right)\nonumber\\
    &\left(\prod_{i=1}^{N_2}\frac{\left(v_1+(K-1)\epsilon_1+K\epsilon_2-u^{(2)}_i\right)\dots\left(v_1-(K-1)\epsilon_1+K\epsilon_2-u^{(2)}_i\right)}{\left(v_2+(K-1)\epsilon_1+K\epsilon_2-u^{(2)}_i\right)\dots\left(v_2-(K-1)\epsilon_1+K\epsilon_2-u^{(2)}_i\right)}\right),
\end{align}
where we have used the equivariant version for simplicity. We conjecture that the equivariant partition function is
\begin{align}
    &Z_\text{mat}=\text{PE}[\mathcal{F}]\left(p'_{(K)},p_{(K,[2,2])},v_1-v_2,\{\epsilon_k\}\right)\nonumber\\
    &\qquad=M\left(p'_{(K)}\right)^{\frac{(v_1-v_2)(\epsilon_1+\epsilon_2)(2\epsilon_1+\epsilon_3)(2\epsilon_1+\epsilon_4)}{2\epsilon_1\epsilon_3\epsilon_4(\epsilon_1-\epsilon_2)}+\frac{(v_1-v_2)\epsilon_3(\epsilon_1-\epsilon_2)(2\epsilon_1-\epsilon_4)}{2\epsilon_1\epsilon_4(\epsilon_1+\epsilon_2)(2\epsilon_1+\epsilon_3)}}\nonumber\\
    &\qquad\quad~\left(M\left(p_{(K,[2,2])},p'_{(K)}\right)\prod_{k=K+1}^{\infty}\frac{1}{\left(1-p_{(K,[2,2])}^{-1}{p'}_{(K)}^k\right)^k}\right)^{\frac{(v_1-v_2)}{\epsilon_4}},\\
    &\mathcal{F}\left(p'_{(K)},p_{(K,[2,2])},v_1-v_2,\{\epsilon_k\}\right)=\left(\frac{(\epsilon_1+\epsilon_2)(2\epsilon_1+\epsilon_3)(2\epsilon_1+\epsilon_4)}{2\epsilon_1\epsilon_3\epsilon_4(\epsilon_1-\epsilon_2)}+\frac{\epsilon_3(\epsilon_1-\epsilon_2)(2\epsilon_1-\epsilon_4)}{2\epsilon_1\epsilon_4(\epsilon_1+\epsilon_2)(2\epsilon_1+\epsilon_3)}\right.\nonumber\\
    &\qquad\qquad\qquad\qquad\qquad\qquad\quad\left.+\frac{p_{(K,[2,2])}+1/p_{(K,[2,2])}}{\epsilon_4}\right)\frac{(v_1-v_2)p'_{(K)}}{\left(1-p'_{(K)}\right)^2}-\frac{1}{p_{(K,[2,2])}\epsilon_4}\sum_{k=1}^Kk{p'}_{(K)}^K\;,
\end{align}
where $p'_{(K)}=-p_1p_2$ and $p_{(K,[2,2])}=-p_1^Kp_2^{K+1}$. In the DT chamber, this agrees with the equivariant DT invariants in \cite{Cao:2019tnw} (upon redefinition of the parameters). The crystals are the 4d versions of \eqref{coninf}, where one ``peels off'' a 3d subcrystal every time one crosses a wall from $C_K$ to $C_{K+1}$.

For the $\widetilde{C}_K$ chamber, the chirals connecting the framing node and the gauge node 2 (resp.~1) have weights $v_2+K\epsilon_1+(K-1)\epsilon_2$, $v_2+(K-2)\epsilon_1+(K-1)\epsilon_2$, $\dots$, $v_2-K\epsilon_1+(K-1)\epsilon_2$ (resp.~$v_1+(K-1)\epsilon_1+(K-1)\epsilon_2$, $v_1+(K-3)\epsilon_1+(K-1)\epsilon_2$, $\dots$, $v_1-(K-1)\epsilon_1+(K-1)\epsilon_2$ when existing) while the accompanied Fermis have weights $v_1+K\epsilon_1+(K-1)\epsilon_2$, $v_1+(K-2)\epsilon_1+(K-1)\epsilon_2$, $\dots$, $v_1-K\epsilon_1+(K-1)\epsilon_2$ (resp.~$-v_2-(K-1)\epsilon_1-(K-1)\epsilon_2$, $-v_2-(K-3)\epsilon_1-(K-1)\epsilon_2$, $\dots$, $-v_2+(K-1)\epsilon_1-(K-1)\epsilon_2$ when existing). Therefore, in the integrand, we only need to change the factors coming from the framing to
\begin{align}
    &\left(\prod_{i=1}^{N_1}\frac{\left(u^{(1)}_i-v_2-(K-1)\epsilon_1-(K-1)\epsilon_2\right)\dots\left(u^{(1)}_i-v_2+(K-1)\epsilon_1-(K-1)\epsilon_2\right)}{\left(u^{(1)}_i-v_1-(K-1)\epsilon_1-(K-1)\epsilon_2\right)\dots\left(u^{(1)}_i-v_1+(K-1)\epsilon_1-(K-1)\epsilon_2\right)}\right)\nonumber\\
    &\left(\prod_{i=1}^{N_2}\frac{\left(v_1+K\epsilon_1+(K-1)\epsilon_2-u^{(2)}_i\right)\dots\left(v_1-K\epsilon_1+(K-1)\epsilon_2-u^{(2)}_i\right)}{\left(v_2+K\epsilon_1+(K-1)\epsilon_2-u^{(2)}_i\right)\dots\left(v_2-K\epsilon_1+(K-1)\epsilon_2-u^{(2)}_i\right)}\right),
\end{align}
where we have used the equivariant version for simplicity. We conjecture that the equivariant partition function is
\begin{align}
    &Z_\text{mat}=\text{PE}[\mathcal{F}]\left(\widetilde{p'}_{(K)},\widetilde{p}_{(K,[2,2])},v_1-v_2,\{\epsilon_k\}\right)=\left(\prod_{k=1}^K\frac{1}{\left(1-\widetilde{p}_{(K,[2,2])}\widetilde{p'}_{(K)}^k\right)^k}\right)^{\frac{(v_1-v_2)}{\epsilon_4}},\\
    &\mathcal{F}\left(\widetilde{p'}_{(K)},\widetilde{p}_{(K,[2,2])},v_1-v_2,\{\epsilon_k\}\right)=\frac{\widetilde{p}_{(K,[2,2])}(v_1-v_2)}{\epsilon_4}\left(\sum_{k=1}^K\sum_{l=1}^k(\sqrt{\mu})^{2l-k-1}\widetilde{p'}_{(K)}^K\right),
\end{align}
where $\widetilde{p'}_{(K)}=-p_1^{-1}p_2^{-1}$ and $\widetilde{p}_{(K,[2,2])}=-p_1^{K+1}p_2^K$. In the PT chamber, this agrees with the rational limit of the K-theoretic PT invariants in \cite{Cao:2019tvv}. The crystals are the 4d versions of \eqref{confin}, where one adds a ``layer'' of a (semi-infinite) 3d subcrystal every time one crosses a wall from $\widetilde{C}_K$ to $\widetilde{C}_{K+1}$. Notice that, however, in contrast to the finite 3d crystals for the conifold, the 4d crystals in the chambers $\widetilde{C}_K$ for the conifold$\times\mathbb{C}$ are infinite (due to the fourth direction extended by the adjoint chirals).

Let us also make a comment on the elliptic invariants. For the chamber $C_K$, the shift of $u^{(1)}_i\rightarrow u^{(1)}_i+a+b\tau$ with $a,b\in\mathbb{Z}$ would yield an extra phase $\text{e}^{-2\pi\text{i}(K+1)(v_1-v_2)}$, and hence $(K+1)(v_1-v_2)\in\mathbb{Z}$. On the other hand, the transformation of $u^{(2)}_i$ would lead to $K(v_1-v_2)\in\mathbb{Z}$. As $K+1$ and $K$ are coprime, we should again have $v_1-v_2\in\mathbb{Z}$. Likewise, for the chamber $\widetilde{C}_K$, the transformations of $u^{(1)}_i$ and $u^{(2)}_i$ would give rise to $K(v_1-v_2)\in\mathbb{Z}$ and $(K+1)(v_1-v_2)\in\mathbb{Z}$. Therefore, we should have $v_1-v_2\in\mathbb{Z}$.

\subsection{\texorpdfstring{Trialities: $Q^{1,1,1}$}{Trialities: Q(1,1,1)}}\label{Q111}
In general, the $\mathcal{N}=(0,2)$ quivers do not need to have 4d counterparts as above. As an example, $Q^{1,1,1}$ has the toric diagram
\begin{equation}
\begin{matrix}
    \includegraphics[width=5cm]{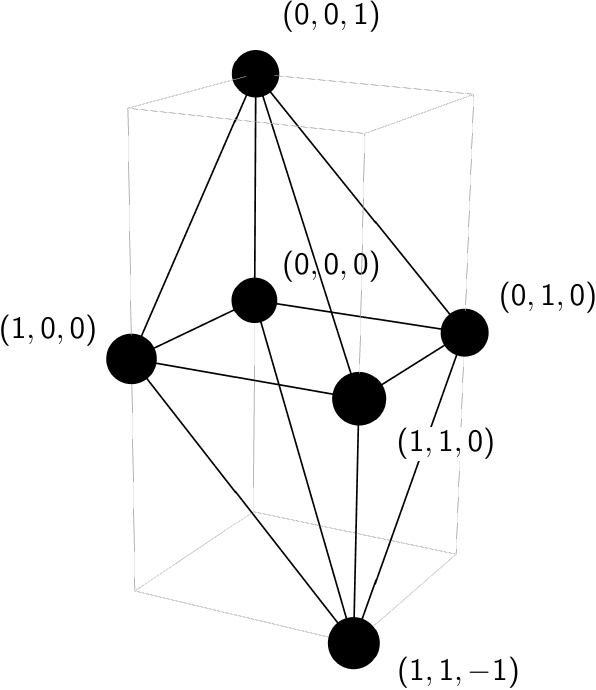}.
\end{matrix}
\end{equation}
In particular, its quiver gauge theories enjoy triality featured in $\mathcal{N}=(0,2)$. Its three phases form the triality network
\begin{equation}
\begin{matrix}
    \includegraphics[width=3cm]{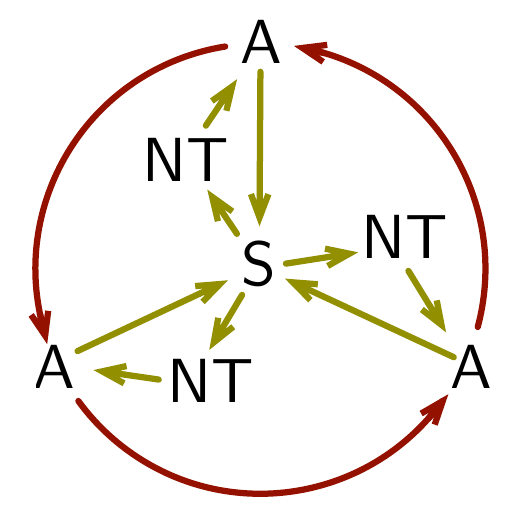}.
\end{matrix}
\end{equation}
Here, phase A (resp.~S) is antisymmetric (resp.~symmetric) in the sense of the permutation of the coordinate axes while phase NT is non-toric. Here, we shall only focus on phase A as an illustration of triality and wall crossing. The outer circle with three quivers of phase A is given by\footnote{The periodic quivers of phase S and phase NT can be found in \cite{Franco:2016nwv}.}
\begin{equation}
\begin{matrix}
    \includegraphics[width=10cm]{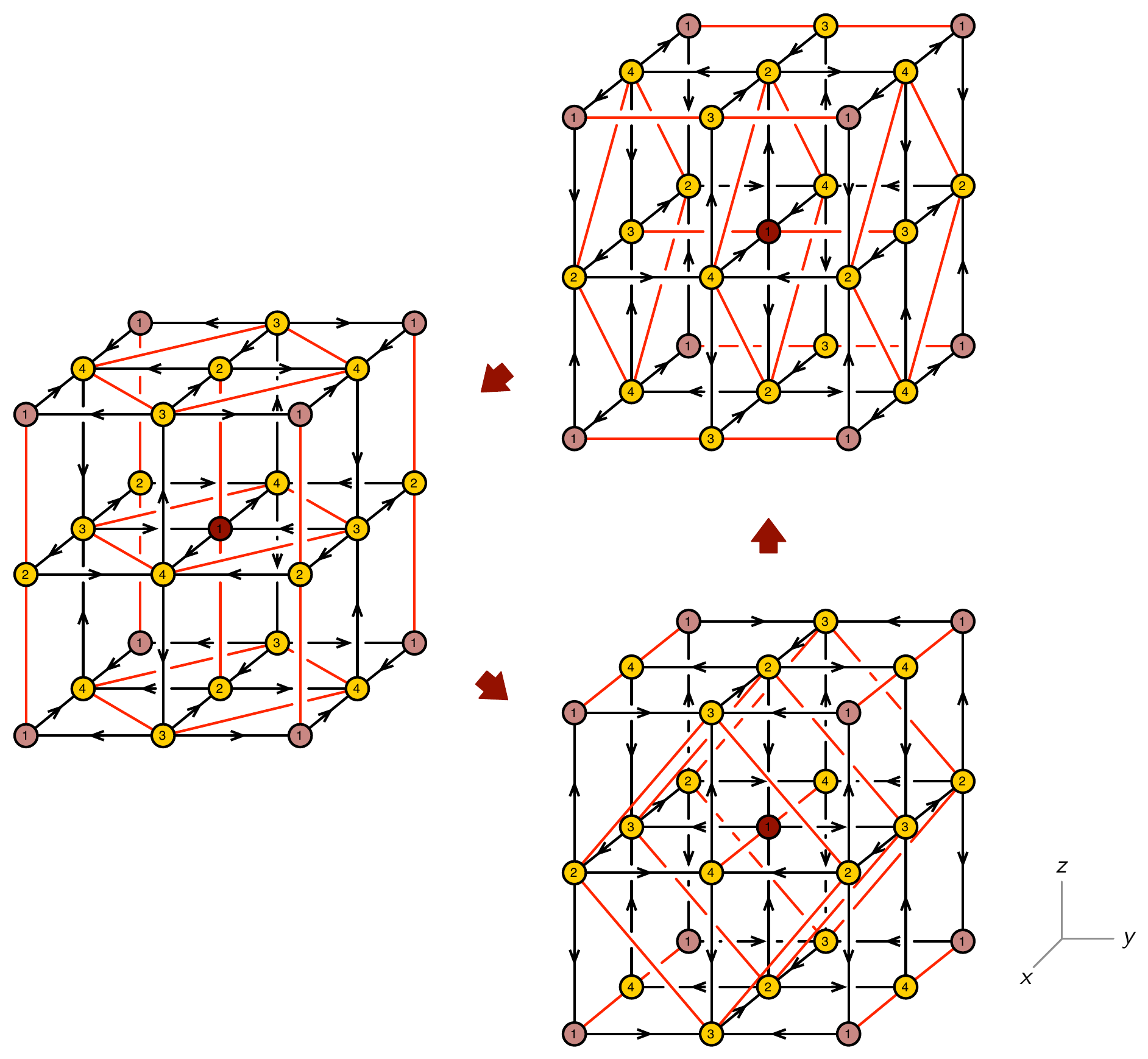}.
\end{matrix}
\end{equation}
This self-triality can be obtained by mutating the node 1 in purple.

Let us start with the quiver
\begin{equation}
\begin{matrix}
    \includegraphics[width=3.5cm]{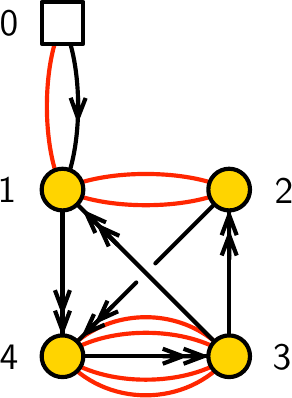}.\label{Q111A1}
\end{matrix}
\end{equation}
The weights of the edges are chosen to be
\begin{equation}
\begin{tabular}{c|c|c|c|c|c|cccc}
$\chi_{14,1}$ & $\chi_{14,2}$ & $\chi_{24,1}$ & $\chi_{24,2}$ & $\chi_{31,1}$ & $\chi_{31,2}$ & \multicolumn{1}{c|}{$\chi_{32,1}$} & \multicolumn{1}{c|}{$\chi_{32,2}$} & \multicolumn{1}{c|}{$\chi_{43,1}$} & $\chi_{43,2}$ \\ \hline
$\epsilon_1$ & $-\epsilon_1$ & $\epsilon_2$ & $-\epsilon_2$ & $\epsilon_2$ & $-\epsilon_2$ & \multicolumn{1}{c|}{$\epsilon_1$} & \multicolumn{1}{c|}{$-\epsilon_1$} & \multicolumn{1}{c|}{$\epsilon_3$} & $-\epsilon_3$ \\ \hline\hline
$\Lambda_{12}$ & $\Lambda_{21}$ & $\Lambda_{34,1}$ & $\Lambda_{34,2}$ & $\Lambda_{34,3}$ & $\Lambda_{34,4}$ & \multicolumn{4}{c}{\multirow{2}{*}{}}                                    \\ \cline{1-6}
$-\epsilon_3$ & $-\epsilon_3$ & $\epsilon_1+\epsilon_2$ & $\epsilon_1+\epsilon_2$ & $\epsilon_1-\epsilon_2$ & $\epsilon_2-\epsilon_1$ & \multicolumn{4}{c}{}
\end{tabular}.
\end{equation}
For the dimension vector $\bm{N}=(N_a)$, the integrand is then
\begin{align}
    Z_\text{1-loop}&=q_3^{N_1N_2+N_3N_4}q_4^{N_3N_4}\left(\prod_{i=1}^{N_1}x^{(1)}_i\right)^{N_1(N_3+N_4-N_2-(N_1-1)/2)}\left(\prod_{i=1}^{N_2}x^{(2)}_i\right)^{N_2(N_3+N_4-N_1-(N_2-1)/2)}\nonumber\\
    &\left(\prod_{i=1}^{N_3}x^{(3)}_i\right)^{N_3(N_1+N_2-N_4-(N_3-1)/2)}\left(\prod_{i=1}^{N_4}x^{(4)}_i\right)^{N_4(N_1+N_2-N_3-(N_4-1)/2)}\left(\sqrt{\frac{w_1}{w_2}}\right)^{N_1}\nonumber\\
    &\left(\prod_{a=1}^4\prod_{i=1}^{N_a}\frac{\text{d}x^{(a)}_i}{x^{(a)}_i}\right)\left(\prod_{i=1}^{N_1}\frac{x^{(1)}_i-w_2}{x^{(1)}_i-w_1}\right)\left(\prod_{a=1}^4\prod_{i\neq j}^{N_a}\left(x^{(a)}_i-x^{(a)}_j\right)\right)\nonumber\\
    &\left(\prod_{i=1}^{N_1}\prod_{j=1}^{N_2}\left(x^{(2)}_j-q_3^{-1}x^{(1)}_i\right)\left(x^{(1)}_i-q_3^{-1}x^{(2)}_j\right)\right)\nonumber\\
    &\left(\prod_{i=1}^{N_1}\prod_{j=1}^{N_3}\frac{1}{\left(x^{(1)}_i-q_2x^{(3)}_j\right)\left(x^{(1)}_i-q_2^{-1}x^{(3)}_j\right)}\right)\left(\prod_{i=1}^{N_1}\prod_{j=1}^{N_4}\frac{1}{\left(x^{(4)}_j-q_1x^{(1)}_i\right)\left(x^{(4)}_j-q_1^{-1}x^{(1)}_i\right)}\right)\nonumber\\
    &\left(\prod_{i=1}^{N_2}\prod_{j=1}^{N_3}\frac{1}{\left(x^{(2)}_i-q_1x^{(3)}_j\right)\left(x^{(2)}_i-q_1^{-1}x^{(3)}_j\right)}\right)\left(\prod_{i=1}^{N_2}\prod_{j=1}^{N_4}\frac{1}{\left(x^{(4)}_j-q_2x^{(2)}_i\right)\left(x^{(4)}_j-q_2^{-1}x^{(2)}_i\right)}\right)\nonumber\\
    &\left(\prod_{i=1}^{N_3}\prod_{j=1}^{N_4}\frac{\left(x^{(4)}_j-q_1q_2x^{(3)}_i\right)^2\left(x^{(4)}_j-q_1^{-1}q_2x^{(3)}_i\right)\left(x^{(4)}_j-q_1q_2^{-1}x^{(3)}_i\right)}{\left(x^{(3)}_i-q_3x^{(4)}_j\right)\left(x^{(3)}_i-q_3^{-1}x^{(4)}_j\right)}\right).
\end{align}
Let us list some indices at low ranks as an illustration:
\begin{itemize}
    \item Level 1:
    \begin{itemize}
        \item crystal labelled by $v_1$ (ranks $(1,0,0,0)$):
        \begin{equation}
            \mathcal{I}_1=\sqrt{\mu}-1/\sqrt{\mu} \;.
        \end{equation}
    \end{itemize}
    \item Level 2:
    \begin{itemize}
        \item crystal labelled by $v_1,v_1+\epsilon_1$ (ranks $(1,0,0,1)$):
        \begin{equation}
            \mathcal{I}_{2,+}=\frac{q_1}{q_1^2-1}\mathcal{I}_1 \;.
        \end{equation}
        \item crystal labelled by $v_1,v_1-\epsilon_1$ (ranks $(1,0,0,1)$):
        \begin{equation}
            \mathcal{I}_{2,-}=\frac{q_1}{1-q_1^2}\mathcal{I}_1 \;.
        \end{equation}
    \end{itemize}
    The index is
    \begin{equation}
        \mathcal{I}_2=\mathcal{I}_{2,+}+\mathcal{I}_{2,-}=0 \;.
    \end{equation}
    \item Level 3:
    \begin{itemize}
        \item crystal labelled by $v_1,v_1+\epsilon_1,v_1-\epsilon_1$ (ranks $(1,0,0,2)$):
        \begin{equation}
            \mathcal{I}_{3,\pm}=w_1^2(\sqrt{\mu}-1/\sqrt{\mu}) \;.
        \end{equation}
        \item crystal labelled by $v_1,v_1+\epsilon_1,v_1+\epsilon_3$ (ranks $(1,0,1,1)$):
        \begin{equation}
            \mathcal{I}_{3,(+,+)}=\frac{q_1q_3q_4(q_1-q_2q_3)(1-q_1q_2q_3)(\sqrt{\mu}-1/\sqrt{\mu})}{(1-q_1^2)(1-q_3^2)} \;.
        \end{equation}
        \item crystal labelled by $v_1,v_1+\epsilon_1,v_1-\epsilon_3$ (ranks $(1,0,1,1)$):
        \begin{equation}
            \mathcal{I}_{3,(+,-)}=\frac{q_1q_3q_4(q_1q_2-q_3)(q_1q_3-q_2)(\sqrt{\mu}-1/\sqrt{\mu})}{(1-q_1^2)(1-q_3^2)} \;.
        \end{equation}
        \item crystal labelled by $v_1,v_1-\epsilon_1,v_1+\epsilon_3$ (ranks $(1,0,1,1)$):
        \begin{equation}
            \mathcal{I}_{3,(-,+)}=\frac{q_1q_3q_4(q_1q_3-q_2)(1-q_1q_2q_3)^2(\sqrt{\mu}-1/\sqrt{\mu})}{(1-q_1^2)(1-q_3^2)(q_1q_2-q_3)}\;.
        \end{equation}
        \item crystal labelled by $v_1,v_1-\epsilon_1,v_1-\epsilon_3$ (ranks $(1,0,1,1)$):
        \begin{equation}
            \mathcal{I}_{3,(-,-)}=\frac{q_1q_3q_4(q_2q_3-q_1)(q_1q_2-q_3)^2(\sqrt{\mu}-1/\sqrt{\mu})}{(1-q_1^2)(1-q_3^2)(1-q_1q_2q_3)}\;.
        \end{equation}
    \end{itemize}
    The indices are
    \begin{align}
        &\mathcal{I}_{(1,0,0,2)}=\mathcal{I}_{3,\pm},\\
        &\mathcal{I}_{(1,0,1,1)}=\mathcal{I}_{3,(+,+)}+\mathcal{I}_{3,(+,-)}+\mathcal{I}_{3,(-,+)}+\mathcal{I}_{3,(-,-)}\nonumber\\
        &\qquad\quad~=\frac{1-4q_1q_2q_3+q_3^2+q_1^2q_2^2(1+q_3^2)}{(q_1q_2-q_3)(q_1q_2q_3-1)}(\sqrt{\mu}-1/\sqrt{\mu})\;.
    \end{align}
\end{itemize}

\paragraph{Wall crossing} We can apply the triality to the quiver and this would take us to a different chamber. Let us mutate the node 1 in \eqref{Q111A1}. This maps to the same quiver but with a relabelling of the nodes by $2\rightarrow3\rightarrow4\rightarrow2$. Taking the framing into account, we have
\begin{equation}
\begin{matrix}
    \includegraphics[trim=0 20 0 0,width=4cm]{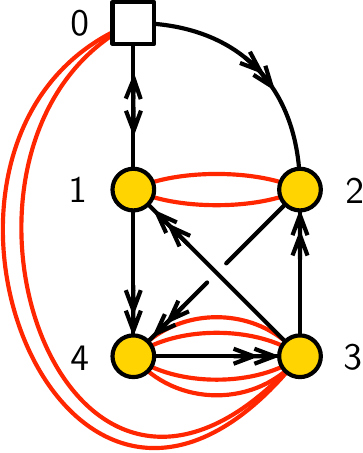}.\label{Q111A2}
\end{matrix}
\end{equation}
The weights of the edges connecting the framing node 0 are taken to be
\begin{equation}
\begin{tabular}{c|c|c|c|c|c}
$\chi_{10}$ & $\chi_{01}$ & $\chi_{02,1}$ & $\chi_{02,2}$ & $\Lambda_{30,1}$ & $\Lambda_{30,2}$ \\ \hline\hline
$v_1$ & $-v_2$ & $v_1+\epsilon_3$ & $-v_1-\epsilon_1-\epsilon_2$ & $v_2+\epsilon_1-\epsilon_3$ & $v_2-\epsilon_2$
\end{tabular}.
\end{equation}
In the integrand, we only need to change the factors coming from the framing to
\begin{align}
    &\left(\prod_{i=1}^{N_1}\frac{1}{\left(v_1-u^{(1)}_i\right)\left(u^{(1)}_i+v_2\right)}\right)\left(\prod_{i=1}^{N_2}\frac{1}{\left(u^{(2)}_i-v_1-\epsilon_3\right)\left(u^{(2)}_i+v_1+\epsilon_1+\epsilon_2\right)}\right)\nonumber\\
    &\left(\prod_{i=1}^{N_3}\left(u^{(3)}_i-v_2-\epsilon_1+\epsilon_3\right)\left(u^{(3)}_i-v_2+\epsilon_2\right)\right),
\end{align}
where we have used the equivariant version for simplicity. Let us list some indices at low ranks as an illustration:
\begin{itemize}
    \item Level 1:
    \begin{itemize}
        \item crystal labelled by $-v_2$ (ranks $(1,0,0,0)$):
        \begin{equation}
            \mathcal{I}_{(1,0,0,0)}=-\frac{w_2}{\sqrt{\mu}-1/\sqrt{\mu}}\;.
        \end{equation}
        \item crystal labelled by $v_1+\epsilon_3$ (ranks $(0,1,0,0)$):
        \begin{equation}
            \mathcal{I}_{(0,1,0,0),1}=\frac{\sqrt{q_1q_2q_3}}{q_3(q_1q_2q_3w_1^2-1)}\;,.
        \end{equation}
        \item crystal labelled by $-v_1-\epsilon_1-\epsilon_2$ (ranks $(0,1,0,0)$):
        \begin{equation}
            \mathcal{I}_{(0,1,0,0),2}=\frac{v_1^2q_1q_2\sqrt{q_1q_2q_3}}{1-q_1q_2q_3v_1^2}\;.
        \end{equation}
    \end{itemize}
    The indices are
    \begin{align}
        &\mathcal{I}_{(1,0,0,0)}=-\frac{w_2}{\sqrt{\mu}-1/\sqrt{\mu}}\;,\\
        &\mathcal{I}_{(0,1,0,0)}=\mathcal{I}_{(0,1,0,0),1}+\mathcal{I}_{(0,1,0,0),2}=-\sqrt{\frac{q_1q_2}{q_3}}\;.
    \end{align}
    \item Level 2:
    \begin{itemize}
        \item crystal labelled by $-v_2,v_1+\epsilon_3$ (ranks $(1,1,0,0)$):
        \begin{equation}
            \mathcal{I}_{(1,1,0,0),1}=\sqrt{\frac{q_1q_2w_2}{q_3w_1}}\frac{(1-w_1w_2)(1-q_3^2w_1w_2)}{q_3(q_1q_2q_3w_1^2-1)(w_1-w_2)}\;.
        \end{equation}
        \item crystal labelled by $-v_2,-v_1-\epsilon_1-\epsilon_2$ (ranks $(1,1,0,0)$):
        \begin{equation}
            \mathcal{I}_{(1,1,0,0),2}=\frac{\sqrt{q_1q_2q_3w_1w_2}w_1(q_1q_2q_3w_1-w_2)(q_3w_2-q_1q_2w_1)}{q_3(q_1q_2q_3w_1^2-1)(w_1-w_2)}\;.
        \end{equation}
        \item crystal labelled by $-v_2,-v_2+\epsilon_1$ (ranks $(1,0,0,1)$):
        \begin{equation}
            \mathcal{I}_{(1,0,0,1),1}=\frac{q_1w_2}{(1-q_1^2)(\sqrt{\mu}-1/\sqrt{\mu})}\;.
        \end{equation}
        \item crystal labelled by $-v_2,-v_2-\epsilon_1$ (ranks $(1,0,0,1)$):
        \begin{equation}
            \mathcal{I}_{(1,0,0,1),2}=\frac{q_1w_2}{(q_1^2-1)(\sqrt{\mu}-1/\sqrt{\mu})}\;.
        \end{equation}
        \item crystal labelled by $v_1+\epsilon_3,v_1+\epsilon_3+\epsilon_2$ (ranks $(0,1,0,1)$):
        \begin{equation}
            \mathcal{I}_{(0,1,0,1),1}=\frac{q_2\sqrt{q_1q_2q_3}}{q_3(q_2^2-1)(q_1q_2q_3w_1^2-1)}\;.
        \end{equation}
        \item crystal labelled by $v_1+\epsilon_3,v_1+\epsilon_3-\epsilon_2$ (ranks $(0,1,0,1)$):
        \begin{equation}
            \mathcal{I}_{(0,1,0,1),2}=\frac{q_2\sqrt{q_1q_2q_3}}{q_3(1-q_2^2)(q_1q_2q_3w_1^2-1)}\;.
        \end{equation}
        \item crystal labelled by $-v_1-\epsilon_1-\epsilon_2,-v_1-\epsilon_1-\epsilon_3+\epsilon_2$ (ranks $(0,1,0,1)$):
        \begin{equation}
            \mathcal{I}_{(0,1,0,1),3}=\frac{q_1q_2^2w_1^2\sqrt{q_1q_2q_3}}{(1-q_2^2)(q_1q_2q_3w_1^2-1)}\;.
        \end{equation}
        \item crystal labelled by $-v_1-\epsilon_1-\epsilon_2,-v_1-\epsilon_1-\epsilon_3-\epsilon_2$ (ranks $(0,1,0,1)$):
        \begin{equation}
            \mathcal{I}_{(0,1,0,1),4}=\frac{q_1q_2^2w_1^2\sqrt{q_1q_2q_3}}{(q_2^2-1)(q_1q_2q_3w_1^2-1)}\;.
        \end{equation}
    \end{itemize}
    The indices are
    \begin{align}
        &\mathcal{I}_{(1,1,0,0)}=\mathcal{I}_{(1,1,0,0),1}+\mathcal{I}_{(1,1,0,0),2}\;,\\
        &\mathcal{I}_{(1,0,0,1)}=\mathcal{I}_{(1,0,0,1),1}+\mathcal{I}_{(1,0,0,1),2}=0\;,\\
        &\mathcal{I}_{(0,1,0,1)}=\mathcal{I}_{(0,1,0,1),1}+\mathcal{I}_{(0,1,0,1),2}+\mathcal{I}_{(0,1,0,1),3}+\mathcal{I}_{(0,1,0,1),4}=0\;.
    \end{align}
\end{itemize}

We may also compare the conditions on the chemical potentials in the elliptic case for the two quivers. For \eqref{Q111A1}, we still have
\begin{equation}
    v_1-v_2\in\mathbb{Z}\;.
\end{equation}
For \eqref{Q111A2}, we have
\begin{equation}
    v_1-v_2\in\mathbb{Z}\;,\quad
    \epsilon_1+\epsilon_2-\epsilon_3\in\mathbb{Z}\;,\quad
    2v_2+\epsilon_1-\epsilon_2-\epsilon_3\in\mathbb{Z}\;.
\end{equation}

\section{Equivariant \texorpdfstring{DT$_4$}{DT4} Invariants}\label{DT4}

It is known that the BPS invariants discussed above are mathematically the (generalized) DT invariants for CY threefolds. In the case of CY fourfolds, there are also extensive studies on counting the coherent sheaves in mathematics literature. Let us now briefly discuss the equivariant DT$_4$ invariants which were established using obstruction theory in \cite{cao2014donaldson,Cao:2017swr,Cao:2019tnw,Cao:2020vce,Cao:2019fqq}. The K-theoretic uplift can be found in \cite{Cao:2019tvv}.

Given a toric CY fourfold $X$, we would like to define some DT(-type) invariants as ``$\int_{[\mathcal{I}(n,\beta)]^\text{vir}}1$'', where $\mathcal{I}(n,\beta)$ is the Hilbert scheme of closed subschemes $Z\subset X$ with $\chi(\mathcal{O}_Z)=n$ and $[Z]=\beta\in H_2(X)$. Here, we will be mainly focusing on the zero-dimensional DT invariants with $\beta=0$, and $\mathcal{I}(n,0)=\text{Hilb}^n(X)$ is the Hilbert scheme of $n$ points on $X$. However, one could quickly run into problems. For example, $\text{Hilb}^n(X)$ is generally non-compact due to the non-compactness of $X$. It turns out that we can consider the fixed locus
\begin{equation}
    \text{Hilb}^n(X)^T=\text{Hilb}^n(X)^{(\mathbb{C}^*)^4}\;,
\end{equation}
where $T\subset(\mathbb{C}^*)^4$ is the 3-dimensional subtorus preserving the CY volume form. This equality and the fact that it consists of finitely many isolated reduced points were shown in \cite{Cao:2017swr,Cao:2019tnw}.

For general $X$, there would be wall crossing, and the BPS indices are expected to related to generalized DT invariants. Therefore, let us first consider the DT chamber. For the $\mathbb{C}^4$ case, each $[Z]$ corresponds to a solid partition\footnote{More generally, $Z$ of dimensions $\leq1$ which can have non-zero $\beta$ were considered in \cite{Cao:2019tnw,Cao:2019tvv}. This further includes the so-called curve-like solid partitions loc.~cit., which are solid partitions with non-trivial asymptotic plane partitions. They are expected to give rise to open BPS states (cf.~\cite{Yamazaki:2010fz,Nagao:2009ky,Galakhov:2021xum}). Although we expect the open BPS states in the cases of general $X$ would also have some combinatorial structures given by the 4d crystals with non-trivial asymptotes, here we shall only focus on the closed BPS states.}. For general $X$, we have shown using the JK residues that they should be labelled by 4d crystals. For each $Z\in\text{Hilb}^n(X)^T$, we have the vector bundles
\begin{equation}
\begin{matrix}
    \includegraphics[width=4cm]{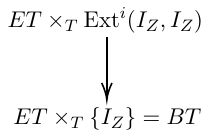}
\end{matrix}
\end{equation}
for $i=1,2$, where $I_Z$ is the monomial ideal that cuts out $Z$,
and $ET \to BT$ is the universal bundle associated with $T$. The deformation and obstruction spaces at $I_Z$ then correspond to $\text{Ext}^1(I_Z,I_Z)$ and $\text{Ext}^2(I_Z,I_Z)$ respectively. Denote their Euler classes as $e_T(\text{Ext}^i(I_Z,I_Z))$. The Serre duality pairing on $\text{Ext}^2(I_Z,I_Z)$ induces a non-degenerate quadratic form $Q$ on the bundle $ET\times_T\text{Ext}^2(I_Z,I_Z)$. This allows one to define the half Euler class
\begin{align}
\begin{split}
    e_T(\text{Ext}^2(I_Z,I_Z),Q)
    &:=\pm\sqrt{(-1)^{\frac{\text{ext}^2(I_Z,I_Z)}{2}}e_T(\text{Ext}^2(I_Z,I_Z))}\\
    &\qquad \in\mathbb{Z}[\epsilon_{1,2,3,4}]/\langle\epsilon_1+\epsilon_2+\epsilon_3+\epsilon_4\rangle\cong\mathbb{Z}[\epsilon_{1,2,3}]\;,
    \end{split}
\end{align}
where the sign is determined by the choice of the orientation on the positive real form of $(ET\times_T\text{Ext}^2(I_Z,I_Z),Q)$. Then the $T$-equivariant virtual fundamental class of $\text{Hilb}^n(X)$ is defined as
\begin{equation}
    [\text{Hilb}^n(X)]_{T,o(\mathcal{L})}^\text{vir}=\sum_{Z\in\text{Hilb}^n(X)^T}\frac{e_T(\text{Ext}^2(I_Z,I_Z),Q)}{e_T(\text{Ext}^1(I_Z,I_Z))}\;,
\end{equation}
where $o\left(\mathcal{L}\right)$ collectively denotes the choice of the square root for each $Z$. Fixing a line bundle $L$ on $X$ with its tautological bundle on $\text{Hilb}^n(X)$ denoted as $L^{[n]}$, the equivariant DT invariant is defined as
\begin{equation}
    \text{DT}_4(n)=\sum_{Z\in\text{Hilb}^n(X)^T}\frac{e_T(\text{Ext}^2(I_Z,I_Z),Q)}{e_T(\text{Ext}^1(I_Z,I_Z))}e_T\left(L^{[n]}\big|_Z\right)\;.
\end{equation}

Recall that in the quiver gauge theory, there is a symmetry between the Fermi multiplets $\Lambda$ and their conjugates $\overline{\Lambda}$, which makes the corresponding edges unoriented in the quiver. In the above mathematical definition, this is reflected by the Serre duality pairing and the quadratic form, which lead to the ``square root'' for $\text{Ext}^2(I_Z,I_Z)$. In other words, the contributions of the Fermi multiplets, which are essentially relations in the gauge theories, are encoded by the obstruction space and the half Euler class $e_T(\text{Ext}^2(I_Z,I_Z),Q)$. On the other hand, the contributions of the chiral multiplets are given by the Euler class of $\text{Ext}^1(I_Z,I_Z)$. Moreover, we have the factor $e_T\left(L^{[n]}\big|_Z\right)$ from the (tautological) insertion. This corresponds to the contribution from the framing, and the line bundle $L$ should be identified with the Chan-Paton bundle on the anti-D8-brane.

In this sense, the BPS counting problem in the fourfold case has some resemblance to the threefold case. Mathematically, it is still determined by the Euler classes of certain $\text{Ext}^1$ and $\text{Ext}^2$ in the deformation-obstruction theory. Physically, certain combinations of the $\mathcal{N}=(0,2)$ multiplets would recover the contributions of the multiplets in the $\mathcal{N}=(2,2)$ case (cf.~Appendix \ref{ellg}).

However, unlike the threefold case, here the Ext$^3$ group does not vanish. We can therefore think of the factor in $[\text{Hilb}^n(X)]^\text{vir}_{T,o(\mathcal{L})}$ as the square root of $\frac{e_T(\text{Ext}^2(I_Z,I_Z))}{e_T(\text{Ext}^1(I_Z,I_Z))e_T(\text{Ext}^3(I_Z,I_Z))}$. The Ext$^1$ and Ext$^2$ groups could be understood as the arrows and plaquettes in the periodic quiver. We expect that the Ext$^3$ group, as the obstructions to the obstructions, corresponds to the relations of the relations. In the 3d periodic quiver, it might be possible that this is reflected by the 3-cells.

Let us also make a comment on the choice of the orientation. Recall that for each Fermi multiplet $\Lambda_i$ (and its conjugate $\overline{\Lambda}_i$), we can make a choice between the $J$-term and the $E$-term. In other words, we can have either $\Lambda_iJ_i$ or $\overline{\Lambda}_iE_i$ in the Lagrangian. As a result, for each crystal configuration, there is a choice of the sign in the 1-loop determinant. This corresponds to a sign choice for each torus fixed point, which is labelled by some 4d crystal, in the mathematical definition. However, in the periodic quiver with its brane brick matching matrix\cite{Franco:2015tya}, once a $J$-/$E$-term is chosen, the others are simultaneously fixed. This would give no ambiguity in the BPS index as the resulting sign is always the same no matter which column one picks from the brick matching matrix. Therefore, the BPS index is uniquely fixed. With this physical input, an orientation is automatically chosen. In some mathematical literature such as \cite{Cao:2017swr,Cao:2019tnw,Cao:2019tvv}, with some specific choices of the orientations (which are conjectured to be unique), the partition functions of some cases were expressed in terms of MacMahon functions and plethystic exponentials. In fact, we observe that such choices always coincide with the physical results.

\paragraph{Non-commutative DT$_4$} The above discussion is restricted to the DT chamber. A mathematical formulation directly defined for the non-commutative DT$_4$ invariants was given in \cite{cao2014donaldson}. As a result, one needs to consider quiver representations rather than sheaves. The Jacobi algebra $A=\mathbb{C}Q/\langle\text{$J$-, $E$-terms}\rangle$, which has the combinatorial structure given by 4d crystals, is a CY$_4$ algebra in the sense that there are pairings $\text{Ext}^i_A(M,N)\times\text{Ext}^{4-i}_A(N,M)\rightarrow\mathbb{C}$ for $A$-modules $M,N$ (with at least one of $M,N$ being finite-dimensional). The representations of the (unframed) quiver (with relations) are in one-to-one correspondence with the finite $A$-modules. Then instead of the Hilbert schemes, one takes the moduli space of framed quiver representations, along with certain framed obstruction bundle endowed with a non-degenerate quadratic form. Then the virtual fundamental class is defined as the Poincar\'e dual of the half Euler class, which lives in the Borel-Moore homology of the moduli space of quiver representations.

\paragraph{Vertex formalism and lifting D8-branes} In \cite{Cao:2017swr,Cao:2019tnw,Cao:2019tvv,Nekrasov:2023nai,Piazzalunga:2023qik}, some vertex formalisms were proposed to obtain the DT invariants. It is tempting to think of this gluing of vertices with certain edge factors as some incarnation of possible ``topological string formalism'' in the fourfold case. Indeed, the subschemes $Z$ are now labelled by a collection of solid partitions, one from each chart in the covering $\{U_{\alpha}\cong\mathbb{C}^4\}$ of $X$.

On the other hand, for CY threefolds (without compact divisors), the connection between BPS partition functions and topological string partition functions via wall crossing was explained from the perspective of M-theory in \cite{Aganagic:2009kf}. There, one puts the M-theory on the Taub-NUT geometry, which is an $S^1$ fibration over $\mathbb{R}^3$ with the circle shrinking at the position of the D6-brane. Then the central charges of the stable BPS particles for different B-fields can be translated to those of the M2-branes.

It is natural to wonder if we can study the (cyclic) wall crossing structures for the CY fourfolds in a similar manner (at least for those without compact 4-/6-cycles). This would also allow us to relate the BPS partition functions to the vertex formalism mentioned above from a more physical perspective. A naive way would be simply replacing the Taub-NUT space with an $S^1$ fibration over $\mathbb{R}$. However, with the presence of the D8-branes, to which the RR field couples, we have the massive Type IIA string theory.
It is believed that massive Type IIA string theory does not admit a strong coupling limit \cite{Aharony:2010af}, and the arguments of \cite{Aganagic:2009kf} does not seem to immediately generalize here. An M9-brane could be reduced to the D8-brane with O8 orientifolds, but it is still not clear whether this would be helpful in the analysis here.

\section*{Acknowledgements}
We would like to thank Ioana Coman, Dmitry Galakhov, Wei Li and Yehao Zhou for enjoyable discussions. JB and MY would like to thank Ulsan National Institute of Science and Technology for hospitality. RKS would like to thank Kavli IPMU for hospitality. 

JB is supported by a JSPS fellowship. RKS is supported by a Basic Research Grant of the National Research Foundation of Korea (NRF2022R1F1A1073128). He is also supported by a Start-up Research Grant for new faculty at UNIST (1.210139.01), a UNIST AI Incubator Grant (1.230038.01) and UNIST UBSI Grants (1.230168.01, 1.230078.01), as well as an Industry Research Project (2.220916.01) funded by Samsung SDS in Korea. He is also partly supported by the BK21 Program (``Next Generation Education Program for Mathematical Sciences'', 4299990414089) funded by
the Ministry of Education in Korea and the National Research Foundation of Korea (NRF). MY is supported in part by the JSPS Grant-in-Aid for Scientific Research (No.~19H00689, 19K03820, 20H05860, 23H01168), and by JST, Japan (PRESTO Grant No.~JPMJPR225A, Moonshot R\&D Grant No.~JPMJMS2061).

\appendix

\section{Elliptic Genera}\label{ellg}

Let us recall the JK residue formulae for the elliptic genus $Z_{T^2}$. The appropriate contour was determined for the rank 1 case in \cite{Benini:2013nda} and was later generalized to any ranks using the recipe of JK residues \cite{Benini:2013xpa}.

Write
\begin{equation}
    x=\text{e}^{2\pi\text{i}u}\;,\quad y=\text{e}^{2\pi\text{i}z}\;,\quad \mathfrak{q}=\text{e}^{2\pi\text{i}\tau}\;.
\end{equation}
For the $\mathcal{N}=(2,2)$ theories, the 1-loop determinant is composed of the following contributions\footnote{Notice that there are some sign differences compared to the ones in \cite{Benini:2013nda,Benini:2013xpa} based on the convention of the fermion number. This is to recover the correct signs in the BPS partition functions for toric CY threefolds as discussed in Appendix \ref{3dcrystals}.}:
\begin{itemize}
    \item Vector multiplet $V$ with gauge group $G$:
    \begin{equation}
        Z_V(\tau,z,u)=\left(-\frac{2\pi\eta(\mathfrak{q})^3}{\theta_1\left(\mathfrak{q},y^{-1}\right)}\right)^{\text{rank}G}\prod_{\alpha\in \Phi(G)}\frac{-\theta_1\left(\mathfrak{q},x^{\alpha}\right)}{\theta_1\left(\mathfrak{q},y^{-1}x^{\alpha}\right)}\prod_{i=1}^{\text{rank}G}\text{d}u_i \;,
    \end{equation}
    where $\Phi$ denotes the root system of $G$ and $x^{\alpha}=\text{e}^{2\pi\text{i}\alpha(u)}$.
    \item Chiral multiplet $\chi$ in the representation $\mathtt{R}$ with $R$-charge $R$:
    \begin{equation}
        Z_{\chi}(\tau,z,u)=\prod_{\rho\in\mathtt{R}}\frac{-\theta_1\left(\mathfrak{q},y^{R/2-1}x^{\rho}\right)}{\theta_1\left(\mathfrak{q},y^{R/2}x^{\rho}\right)}\;.
    \end{equation}
    \item Twisted chiral multiplet $\Sigma$ with axial $R$-charge $R_A$:
    \begin{equation}
        Z_{\Sigma}(\tau,z)=\frac{-\theta_1\left(\mathfrak{q},y^{-R_A/2+1}\right)}{\theta_1\left(\mathfrak{q},y^{-R_A/2}\right)}\;.
    \end{equation}
\end{itemize}
The expressions of the elliptic functions are given below.

For the $\mathcal{N}=(0,2)$ theories, the 1-loop determinant consists of the following contributions:
\begin{itemize}
    \item Vector multiplet:
    \begin{equation}
        Z_V(\tau,u)=\left(\frac{2\pi\eta(\mathfrak{q})^2}{\text{i}}\right)^{\text{rank}G}\prod_{\alpha\in \Phi(G)}\text{i}\frac{\theta_1\left(\mathfrak{q},x^{\alpha}\right)}{\eta\left(\mathfrak{q}\right)}\prod_{i=1}^{\text{rank}G}\text{d}u_i\;.
    \end{equation}
    \item Chiral multiplet:
    \begin{equation}
        Z_{\chi}(\tau,u)=\prod_{\rho\in\mathtt{R}}\text{i}\frac{\eta\left(\mathfrak{q}\right)}{\theta_1\left(\mathfrak{q},x^{\rho}\right)}\;.
    \end{equation}
    \item Fermi multiplet $\Lambda$:
    \begin{equation}
        Z_{\Lambda}(\tau,u)=\prod_{\rho\in\mathtt{R}}\text{i}\frac{\theta_1\left(\mathfrak{q},x^{\rho}\right)}{\eta\left(\mathfrak{q}\right)}\;.
    \end{equation}
\end{itemize}

We have used the Dedekind eta function and a Jacobi theta function above:
\begin{equation}
    \eta(\mathfrak{q})=\mathfrak{q}^{1/24}\prod_{k=1}^{\infty}\left(1-\mathfrak{q}^k\right)\;,
    \quad\theta_1(\tau,z)=-\text{i}\mathfrak{q}^{1/8}y^{1/2}\prod_{k=1}\left(1-\mathfrak{q}^k\right)\left(1-y\mathfrak{q}^k\right)\left(1-y^{-1}\mathfrak{q}^{k-1}\right)\;.
\end{equation}
For $a,b\in\mathbb{Z}$, we have the transformation
\begin{equation}
    \theta_1(\tau,u+a+b\tau)=(-1)^{a+b}\text{e}^{-2\pi\text{i}bu-\pi\text{i}b^2\tau}\theta_1(\tau,u)\;.
\end{equation}
In this paper, we are mainly focusing on the Witten index, which can be obtained from dimensional reduction in the limit $\mathfrak{q}\rightarrow0$. As $(\mathfrak{q};\mathfrak{q})_{\infty}=1$ and $(y;\mathfrak{q})_{\infty}=1-y$ in this limit, we have
\begin{equation}
    \eta(\tau)=\mathfrak{q}^{1/24}\;,
    \quad\theta_1(\tau,z)=\text{i}\mathfrak{q}^{1/8}y^{-1/2}(1-y)=2\mathfrak{q}^{1/8}\sin(\pi z)
    \;.
\end{equation}

\section{Derivation of 3d Crystals from JK Residues}\label{3dcrystals}

It is well-known that the BPS counting problem in the toric CY threefold setting is encoded by 3d crystals \cite{Okounkov:2003sp,Ooguri:2008yb}. We will now show this using the JK residue technique. The strategy is the same as the one for the 4d crystal case. The contribution from the vector multiplet is given by
\begin{equation}
    Z_V=\left(\frac{1}{y^{1/2}\left(1-y^{-1}\right)}\right)^{\text{rank}G}\left(\prod_{\alpha\in \Phi(G)}\frac{-\left(1-x^{\alpha}\right)}{y^{1/2}\left(1-y^{-1}x^{\alpha}\right)}\right)\prod_{i=1}^{\text{rank}G}\frac{\text{d}x_i}{x_i}\;,
\end{equation}
while the contribution from the chiral multiplet is
\begin{equation}
    Z_{\chi}=\prod_{\rho\in\mathtt{R}}\frac{-y\left(1-y^{R/2-1}x^{\rho}\right)}{\left(1-y^{R/2}x^{\rho}\right)}\;.
\end{equation}
Using the weights $q_i$ such that $q:=q_1q_2q_3=y^{-1}$, we have
\begin{align}
    \Delta Z_{N-1,N}=&-\frac{q^{1/2}}{1-q}\left(\prod_{i=1}^{N_a-1}q^{1/2}\frac{\left(1-x^{(a)}_{N_a}\left(x^{(a)}_i\right)^{-1}\right)\left(1-x^{(a)}_i\left(x^{(a)}_{N_a}\right)^{-1}\right)}{\left(1-qx^{(a)}_{N_a}\left(x^{(a)}_i\right)^{-1}\right)\left(1-qx^{(a)}_i\left(x^{(a)}_{N_a}\right)^{-1}\right)}\right)\nonumber\\
    &\left(\prod_{b\in Q_0}\prod_{i=1}^{N_b}\left(\prod_{I\in\{a\rightarrow b\}}q^{-1}\frac{1-qq_I^{-1}x^{(b)}_i\left(x^{(a)}_{N_a}\right)^{-1}}{1-q_I^{-1}x^{(b)}_i\left(x^{(a)}_{N_a}\right)^{-1}}\right)\left(\prod_{I\in\{b\rightarrow a\}}q^{-1}\frac{1-qq_I^{-1}x^{(a)}_{N_a}\left(x^{(b)}_i\right)^{-1}}{1-q_I^{-1}x^{(a)}_{N_a}\left(x^{(b)}_i\right)^{-1}}\right)\right)\nonumber\\
    &\left(\prod_{I\in\{a\rightarrow a\}}q^{-1}\frac{1-qq_I^{-1}}{1-q_I^{-1}}\right)\left(q^{-1}\frac{1-qw^{-1}x^{(a)}_{N_a}}{1-w^{-1}x^{(a)}_{N_a}}\right)^{\delta_{a,0}}\frac{\text{d}x^{(a)}_{N_a}}{x^{(a)}_{N_a}}\;,
\end{align}
where $w=\text{e}^{2\pi\text{i}v}$ denotes the weight of the arrow from the framing node to the initial node (labelled by 0). The crystal structure can be seen in a similar manner as the 4d crystal case discussed in the main context, where the cancellations of the unwanted poles by the contributions from the Fermi multiplets are replaced by the contributions from the chirals pointing backwards in the crystal here.

As all the factors from (the roots of) the vector multiplets and the chirals are in pairs in the numerator and the denominator, there would be one factor in the numerator left are taking the residue (which cancels the factor $x^{(a)}_{N_a}$ in the denominator). This factor in the numerator would then be cancelled by the factor $(1-q)$ in the denominator. More concretely, suppose that we take the pole at $\left(1-\mathfrak{f}x^{(a)}_{N_a}\right)$ for some factor $\mathfrak{f}$. The residue of $1/\left(1-\mathfrak{f}x^{(a)}_{N_a}\right)$ at $x^{(a)}_{N_a}=\mathfrak{f}^{-1}$ is $-\mathfrak{f}^{-1}$, which cancels the factor $x^{(a)}_{N_a}$ in the expression with a minus sign left. The corresponding numerator of this pole $\left(1-q\mathfrak{f}x^{(a)}_{N_a}\right)$ is then cancelled by $1/(1-q)$ in the expression. After taking $q=1$, all the paired factors in the numerator and the denominator get cancelled, and $\Delta Z_{N-1,N}$ is simply $\pm1$.

Let us now determine the sign factor\footnote{If we use the convention of \cite{Benini:2013nda,Benini:2013xpa,Cordova:2014oxa}, $\Delta Z_{N-1,N}$ would always be $+1$, and this would recover the generating functions of the 3d crystals (whose coefficients are always positive).}. It is not hard to see that $Z_N$ has the sign
\begin{equation}
    (-1)^{d_1+\langle\bm{d},\bm{d}\rangle}\;.
\end{equation}
Here, $d_1$ denotes the rank of the gauge node connected to the framing node, and
\begin{equation}
    \langle\alpha,\beta\rangle=\sum_{a\in Q_0}\alpha_a\beta_a-\sum_{\{a\rightarrow b\}}\alpha_a\beta_b
\end{equation}
is the Ringel form with the dimension vector $\bm{d}=\left(d_a\right)$ such that $\sum\limits_{a\in Q_0}d_a=N$. This is precisely the sign factor for the BPS index as given in \cite{mozgovoy2010noncommutative}.

Without taking $q=1$, this gives a refinement of the indices\footnote{Some examples of the equivariant version were studied in \cite{zhou2015donaldson}.}. We expect that this would also recover the refined BPS indices that further track the spin information, as was discussed in \cite{Beaujard:2019pkn,Descombes:2021snc}\footnote{We are grateful to Pierre Descombes and Boris Pioline for insightful comments on this point.}.


\addcontentsline{toc}{section}{References}
\bibliographystyle{utphys}
\bibliography{references}

\end{document}